\documentclass[prd,reprint,superscriptaddress,altaffillsymbol,amsmath,nofootinbib]{revtex4-1}
\pdfoutput=1

\usepackage{ifthen}
\usepackage{epsfig}
\usepackage{booktabs} 
\usepackage{natbib}
\usepackage{wasysym}
\usepackage{feynmp}
\usepackage{slashed} 
\usepackage{bbm}
\usepackage{mathrsfs}
\usepackage{amssymb}
\usepackage{dsfont}
\usepackage[export]{adjustbox}

\newcommand{\boldmidrule}{\specialrule{1.5pt}{0em}{0em}}

\usepackage{color}

\newcommand{\beqra}{\begin{eqnarray}}
\newcommand{\eeqra}{\end{eqnarray}}
\newcommand{\beq}{\begin{equation}}
\newcommand{\eeq}{\end{equation}}

\newcommand{\unit}[1]{\,\mathrm{#1}}
\newcommand{\gev}{\unit{GeV}}
\newcommand{\tev}{\unit{TeV}}
\newcommand{\fb}{\unit{fb}}

\newcommand{\ii}{\mathrm{i}}

\renewcommand{\epsilon}{\varepsilon}
\newcommand{\gDM}{g_\text{DM}}
\newcommand{\ETmiss}{\slashed{E}_\text{T}}

\newcommand{\refeq}[1]{Eq.~(\ref{#1})}
\newcommand{\reftab}[1]{Tab.~\ref{#1}}
\newcommand{\sect}{Sec.}

\newcommand{\Ref}{Ref.}
\newcommand{\Refs}{Refs.}
\newcommand{\app}{Appendix}
\renewcommand{\vec}[1]{\mathbf{#1}}
\renewcommand{\bar}{\overline}

\begin{document}

\title{Determining Dark Matter properties with a XENONnT/LZ signal and LHC-Run3 mono-jet searches}

\author{Sebastian Baum}
\email{sbaum@fysik.su.se}
\affiliation{Oskar Klein Centre, Department of Physics, Stockholm University, AlbaNova, Stockholm SE-10691, Sweden}

\author{Riccardo Catena}
\email{catena@chalmers.se}
\affiliation{Chalmers University of Technology, Department of Physics, SE-412 96 G\"oteborg, Sweden}

\author{Jan Conrad}
\email{conrad@fysik.su.se}
\affiliation{Oskar Klein Centre, Department of Physics, Stockholm University, AlbaNova, Stockholm SE-10691, Sweden}

\author{Katherine Freese}
\email{katie.freese@fysik.su.se}
\affiliation{Oskar Klein Centre, Department of Physics, Stockholm University, AlbaNova, Stockholm SE-10691, Sweden}
\affiliation{Department of Physics, University of Michigan, Ann Arbor, MI 48109, US}

\author{Martin B.~Krauss}
\email{martin.krauss@chalmers.se}
\affiliation{Chalmers University of Technology, Department of Physics, SE-412 96 G\"oteborg, Sweden}

\begin{abstract} 
We develop a method to forecast the outcome of the LHC Run 3 based on the hypothetical detection of $\mathcal{O}(100)$ signal events at XENONnT.~Our method relies on a systematic classification of renormalisable single-mediator models for dark matter-quark interactions, and is valid for dark matter candidates of spin less than or equal to one.~Applying our method to simulated data, we find that at the end of the LHC Run 3 only two mutually exclusive scenarios would be compatible with the detection of $\mathcal{O}(100)$ signal events at XENONnT.~In a first scenario, the energy distribution of the signal events is featureless, as for canonical spin-independent interactions.~In this case, if a mono-jet signal is detected at the LHC, dark matter must have spin 1/2 and interact with nucleons through a unique velocity-dependent operator.~If a mono-jet signal is not detected, dark matter interacts with nucleons through canonical spin-independent interactions.~In a second scenario, the spectral distribution of the signal events exhibits a bump at non zero recoil energies.~In this second case, a mono-jet signal can be detected at the LHC Run 3, dark matter must have spin 1/2 and interact with nucleons through a unique momentum-dependent operator.~We therefore conclude that the observation of $\mathcal{O}(100)$ signal events at XENONnT combined with the detection, or the lack of detection, of a mono-jet signal at the LHC Run 3 would significantly narrow the range of possible dark matter-nucleon interactions.~As we argued above, it can also provide key information on the dark matter particle spin.
\end{abstract}

\maketitle

\section{Introduction}
\label{sec:introduction}
Compelling evidence shows that significant amounts of dark matter (DM) are present in the Universe~\cite{Bertone:2016nfn}.~This conclusion is supported by increasingly accurate cosmological and astrophysical observations performed on extremely different physical scales, ranging from the solar neighbourhood~\cite{Iocco:2015xga} up to the largest scales we can probe in the Cosmos~\cite{Ade:2015xua}.~This picture is further consolidated by the large number of theoretical models which naturally predict DM candidates that are in good agreement with experimental constraints on physics beyond the Standard Model~\cite{Bertone:2004pz}.~In recent years, the search for DM has been successfully pushed forward on several frontiers, and further progress in this area is already foreseeable in the near future with planned and upcoming experiments.~In particular, the search for nuclear recoil events induced by the non-relativistic scattering of Milky Way DM particles in low-background detectors -- a technique known as direct detection~\cite{Drukier:1983gj,Goodman:1984dc,Drukier:1986tm} -- and the search for DM particles produced in energetic collisions at particle accelerators such as the Large Hadron Collider (LHC) have proven to be very effective and highly complementary approaches to explore the vast range of possibilities in the space of DM particle theories~\cite{Bergstrom:2000pn}.

The physics reach of direct detection experiments has improved dramatically in recent years.~The recently released first XENON1T data set the most stringent exclusion limits on the spin-independent DM-nucleon scattering cross-section for DM masses below 100 GeV, with a minimum of $7.7\times10^{-47}$~cm$^{2}$ for a 35 GeV DM particle at 90\% confidence level~\cite{Aprile:2017iyp}.~For DM particle masses above 100 GeV, the most stringent exclusion limits have recently been reported by the PANDAX-II experiment~\cite{Cui:2017nnn}.~The results in Refs.~\cite{Aprile:2017iyp,Cui:2017nnn} improve previous limits published by the LUX collaboration~\cite{Akerib:2016vxi}.~Significantly smaller scattering cross-sections will be probed by next generation experiments such as XENONnT~\cite{Aprile:2015uzo}, LZ~\cite{Mount:2017qzi} and DARWIN~\cite{Aalbers:2016jon} from $\sim 2019$ onwards, with great expectations for ground-breaking discoveries.~Based on this progress and expectations, the first direct detection of DM particles is a concrete prospect for the next few years, especially if DM is made of Weakly Interacting Massive Particles (WIMPs)~\cite{Roszkowski:2017nbc,Bertone:2010at}.~At the same time, the search for new physics beyond the Standard Model and particle DM at the LHC is progressing rapidly.~The LHC Run 2 has started in 2015 producing the first collisions at the centre-of-mass energy $\sqrt{s} = 13 \tev$.~After a long shutdown period between 2018 and 2019, the Run 3 phase will start in 2020, reaching the expected integrated luminosity of 300 fb$^{-1}$ in 2022~\cite{Kahlhoefer:2017dnp}.~A variety of processes will be scrutinised by the ATLAS and CMS collaborations in the data analysis.~Crucial for the present work is the search for DM particles in final states including a single jet plus missing transverse energy.

Combining results from direct detection experiments and DM searches at the LHC is known to be crucial to narrow the allowed regions in the DM particle theory space.~This consideration applies to the interpretation of current null results, but will also apply to the analysis of hypothetical positive results in the future.~Different approaches have been developed for this purpose in the past few years.~In a first approach, results from direct detection experiments and the LHC are combined in so-called global fits, e.g.~\cite{Workgroup:2017lvb,Roszkowski:2017dou,Roszkowski:2016bhs}, where theoretical predictions are expressed in terms of common model parameters, and all experimental information is encoded in a single likelihood  function.~In a second approach -- only applicable in the presence of a signal at the LHC -- theoretical predictions, e.g.,~based upon low-energy supersymmetric theories, are fitted to a sample of simulated LHC data -- typically spectroscopic measurements~\cite{Baltz:2006fm,Bertone:2010rv}.~Through this fit, predictions for WIMP properties such as relic density, annihilation and scattering cross-sections can be extracted.~Recently, simulated LHC dileptonic events have also been analysed to reveal properties of DM, such as the spin~\cite{Capdevilla:2017doz}.~Proposals for extracting the DM particle spin and particle-antiparticle nature from direct detection data can be found in~\cite{Queiroz:2016sxf,Kavanagh:2017hcl,Catena:2017wzu}.

In this work, we propose a third approach to combine direct detection experiments and DM searches at the LHC.~Specifically, the purpose of this work is to quantitatively answer the following question:~assuming $\mathcal{O}(100)$ signal events at XENONnT with an exposure of 20~ton$\times$year, as if the strength of DM-nucleon interactions were just below current exclusion limits, what concrete predictions can be made for the outcome of the mono-jet searches that will be performed at the LHC at the end of Run 3?~Answering this question is crucial:~because of the timeline reviewed above and of the recent progress in the field of DM direct detection, the first unambiguous signal of particle DM might realistically be observed at next generation direct detection experiments, such as XENONnT.~We will address this question within the theoretical framework recently introduced in~\cite{Dent:2015zpa}, where DM-quark interactions are modelled in terms of simplified models.~This framework consists of the most general set of renormalisable models compatible with Lorentz- and gauge-symmetry extending the Standard Model by one DM candidate and one particle mediating DM-quark interactions.~Within this framework, we will develop a method to forecast the outcome of the LHC Run 3 based upon the hypothetical detection of $\mathcal{O}(100)$ signal events at XENONnT.~Our method consists of two stages:

\begin{enumerate}
\item In a first stage, we identify the mono-jet production cross-sections that are compatible with the observation of $\mathcal{O}(100)$ signal events at XENONnT for each simplified model in~\cite{Dent:2015zpa}.~More specifically, for each simplified model $\mathscr{M}$ in~\cite{Dent:2015zpa}, we show that $\mathcal{O}(100)$ signal events observed at XENONnT can only be explained in a relatively narrow region $\mathscr{S}$ of the $M_{\rm med}-\sigma$ plane, where $M_{\rm med}$ is the mass of the particle that mediates the DM interactions with quarks and $\sigma$ is the cross-section for mono-jet production via proton-proton collision.~If DM is described by $\mathscr{M}$, the LHC searches for mono-jet events should focus on $\mathscr{S}$.

\item In a second stage, we identify the correct (family of) DM-nucleon interaction(s) and possibly the DM particle spin based upon the observation, or the lack of observation, of a mono-jet signal at the end of the LHC Run 3.~Importantly, the only input needed here is the knowledge of whether a mono-jet signal has been observed or not.~Further information on the outcome of the LHC Run 3 follows as an output from the application of our method.~The second stage of our method relies on the following considerations.~Constructing $\mathscr{S}$ does not require information on the energy spectrum of the $\mathcal{O}(100)$ signal events observed at XENONnT.~In general, the simplified models in~\cite{Dent:2015zpa} predict nuclear recoil energy spectra which can be divided into two classes, here labelled by $A$ and $B$.~Spectra of type $A$ have a maximum at $q=0$, where $q$ is the transferred momentum.~For example, canonical spin-independent interactions generate spectra of this type.~Spectra of type $B$ have a maximum at $q\neq0$.~Here we will demonstrate that $\mathcal{O}(100)$ signal events at XENONnT will be enough to statistically discriminate spectra of type $A$ from spectra of type $B$ if energy information is used.~Consequently, some of the models in~\cite{Dent:2015zpa} can be rejected based on XENONnT data alone.~This observation will further restrict the regions in the $M_{\rm med}-\sigma$ plane where a mono-jet signal is expected at the LHC Run 3.
\end{enumerate}
We will apply our method to simulated XENONnT data.~In so doing, we will show that at the end of the LHC Run 3 only two mutually exclusive scenarios will be compatible with the observation of $\mathcal{O}(100)$ signal events at XENONnT:
\begin{enumerate}
\item XENONnT detects $\mathcal{O}(100)$ signal events with a spectral distribution of type $A$.~If a mono-jet signal is detected at the LHC, DM must have spin 1/2 and its interactions with nucleons must be of type $\hat{\mathcal{O}}_8$, using the notation introduced in Sec.~\ref{sec:theory}.~If a mono-jet signal is not detected, DM-nucleon interactions must be of type $\hat{\mathcal{O}}_1$.
\item  XENONnT detects $\mathcal{O}(100)$ signal events with a spectral distribution of type $B$.~In this case, a mono-jet signal can be detected, DM must have spin 1/2 and interact with nucleons through the operator $\hat{\mathcal{O}}_{11}$.
\end{enumerate}
These considerations show that model-independent predictions for the LHC Run 3 based upon direct detection data only are indeed possible.~At the same time, our conclusions rely on the framework introduced in~\cite{Dent:2015zpa}, neglect operator evolution and chiral effective field theory corrections to the DM couplings to protons and neutrons, do not consider DM-quark interactions mediated by particles that are charged under the Standard Model gauge group, and focus on elastic DM-nucleus scattering.~Furthermore, we do not impose constraints from thermal production of DM.

This paper is structured as follows.~In \sect~\ref{sec:theory} we review the theoretical framework introduced in~\cite{Dent:2015zpa} to model DM-quark interactions and their non-relativistic limit.~We then discuss the characteristics of a hypothetical XENONnT signal within this framework in \sect~\ref{sec:detection}.~Assuming the observation of such a signal, we will comment on the implications for the LHC mono-jet searches in \sect~\ref{sec:LHC}.~We will finally comment on the validity regime of our results in \sect~\ref{sec:discussion} and conclude in Sec.~\ref{sec:conclusion}.

\section{Theoretical Framework}
\label{sec:theory}

\subsection{Simplified models for DM-quark interactions}
A variety of theories extending beyond the Standard Model accommodate candidates for particle DM.~Even minimal and constrained versions of these, usually can only be studied in detail in certain limiting cases or under simplifying assumptions.~Since the complex details of such theories can, however, often be neglected in the analysis of processes involving DM, a more model independent approach might give more insights and be easier to handle.~This is where the concept of ``simplified models'' can be applied~\cite{Abdallah:2014hon,Buchmueller:2013dya}.~In simplified models, the Standard Model is only extended by the DM itself and an additional new particle that mediates interactions between the SM particles and the dark sector.~This approach is usually sufficient to obtain a good understanding of the collider phenomenology of the DM particle.~It can also provide a better description of processes involving the mediator particle, as opposed to a pure effective theory approach.~Consequently, ATLAS and CMS have increasingly made use of simplified models in their analysis in recent years, e.g.,~\cite{Abercrombie:2015wmb}.~Simplified models can generally be characterised by the nature of the DM particle, i.e., scalar $S$, fermionic $\chi$ or vector DM $X^\mu$.~The DM particle must be neutral with respect to the Standard Model gauge group.~It can, however, carry charge under some additional gauge group.~For our purposes, it is sufficient to assume that the only additional symmetry is a parity under which DM is odd and all other particles are even.~This parity guarantees that DM is stable on cosmological time-scales.~As a consequence, $S$ and $X^\mu$ must be complex fields.~For the case of spin 1/2 DM, we will assume that $\chi$ is a Dirac rather than Majorana fermion. For instance, the Lagrangian of a simplified model with fermionic DM $\chi$ and a vector mediator $G_\mu$, e.g.~a $Z'$-boson, would read as follows
\begin{align}
 \mathcal{L} \ &= \mathcal{L}_\text{SM} + \ii \bar\chi \slashed{D}\chi - m_\chi \bar \chi \chi \nonumber \\ &
 -\frac{1}{4} \mathcal{G}_{\mu\nu}\mathcal{G}^{\mu\nu} + \frac{1}{2} M_{G}^2 G_\mu {G}^\mu \nonumber\\ &
 + \ii \bar q \slashed{D} q - m_q \bar q q \nonumber\\ &
 - \lambda_3 \bar \chi \gamma^\mu \chi G_\mu - \lambda_4 \bar \chi \gamma^\mu \gamma^5 \chi G_\mu \nonumber\\ &
 - h_3 \bar q \gamma^\mu q G_\mu - h_4 \bar q \gamma^\mu \gamma^5 q G_\mu \,,
 \label{eq:LagrGchi}
\end{align}
where the $\lambda_i$ and $h_i$ are dimensionless couplings, and terms involving quark bilinears should be understood as summing over all quark flavors $(q=u,d,c,s,b,t)$, e.g. $h_3 \bar q \gamma^\mu q G_\mu \equiv \sum_{q} h_3^q \bar q \gamma^\mu q G_\mu $.~For this work, we consider universal quark-mediator couplings, $h_i^q = h_i$.~The Lagrangians for all models that we consider in this study can be found in \Ref~\cite{Dent:2015zpa}.~For the convenience of the reader we list them in \app~\ref{app:Lagrangians}.~We can distinguish these models further depending on which couplings in the Lagrangians are non-zero.~E.g., in Eq.~(\ref{eq:LagrGchi}) the case $\lambda_3,h_3 \neq 0$ would correspond to a vector mediator whereas the case $\lambda_4,h_4 \neq 0$ corresponds to an axial-vector mediator.~We will not consider cases where more than two couplings are different from zero at the same time. 

\subsection{Non-relativistic DM-nucleon interactions}
In the non-relativistic scattering of Milky Way DM particles by nuclei, typical momentum transfers are below $200$~MeV.~If the particle mediating the interactions between DM and quarks is significantly heavier than this energy scale, it can be eliminated from the mass spectrum of the theory and its contribution to the scattering process be encoded in higher dimensional operators describing contact interactions between DM and quarks, i.e.~it can be integrated out.~This leads to an effective theory description of DM-quark and -nucleon interactions.~A systematic classification of the interaction operators that can arise from the non-relativistic reduction of simplified models has initially been proposed in~\Ref~\cite{Fitzpatrick:2012ix}.~These interaction operators must be invariant under Galilean transformations\footnote{Galilean invariance can be broken by sub-leading interaction operators~\cite{Bishara:2017pfq}.} -- constant shifts of particle velocities -- and Hermitian conjugation.~Their matrix elements between incoming and outgoing DM-nucleon states can be expressed in terms of the basic invariants under these symmetries
\begin{equation}
 \ii \vec q\,,\quad
 \vec v^\perp \equiv \vec v + \frac{\vec q}{2 \mu_N}\,,\quad
 \vec S_\chi\,,\quad\text{and} \quad
 \vec S_N\,,
\end{equation}
where $\vec q$ is the three-dimensional momentum transferred in the elastic scattering, $\vec v$ is the DM-nucleon relative velocity, and $\vec S_{\chi}$ ($\vec S_N$) is the spin of the DM (nucleon), respectively.~The classification initially proposed in~\cite{Fitzpatrick:2012ix} for spin $\le 1/2$ DM has subsequently been extended to spin 1 DM in~\cite{Dent:2015zpa}.~In particular, it has been shown that 16 independent operators can be constructed at linear order in the transverse relative velocity $ \vec v^\perp$, although not all of them appear as leading operators in the non-relativistic limit of the simplified models.~The 16 operators for DM-nucleon interactions identified in~\cite{Dent:2015zpa,Fitzpatrick:2012ix} are listed in \reftab{tab:operators}.~Operators $\hat{\mathcal{O}}_{17}$ and $\hat{\mathcal{O}}_{18}$ only arise for spin 1 DM~\cite{Dent:2015zpa}.

\begin{table}[t!]
    \centering
    \begin{ruledtabular}	
    \begin{tabular}{lc}
    \toprule
    \boldmidrule
    	Operator & Type of spectrum \\
	 \toprule
	  \toprule
        $\hat{\mathcal{O}}_1 = \mathds{1}_{\chi}\mathds{1}_N$  & $A$\\  
        $\hat{\mathcal{O}}_3 = i{\bf{\hat{S}}}_N\cdot\left(\frac{{\bf{\hat{q}}}}{m_N}\times{\bf{\hat{v}}}^{\perp}\right)\mathds{1}_\chi$ & $B$\\
        $\hat{\mathcal{O}}_4 = {\bf{\hat{S}}}_{\chi}\cdot {\bf{\hat{S}}}_{N}$ & $A$ \\                                                      
        $\hat{\mathcal{O}}_5 = i{\bf{\hat{S}}}_\chi\cdot\left(\frac{{\bf{\hat{q}}}}{m_N}\times{\bf{\hat{v}}}^{\perp}\right)\mathds{1}_N$ & $B$\\       
        $\hat{\mathcal{O}}_6 = \left({\bf{\hat{S}}}_\chi\cdot\frac{{\bf{\hat{q}}}}{m_N}\right) \left({\bf{\hat{S}}}_N\cdot\frac{\hat{{\bf{q}}}}{m_N}\right)$& $B$\\  
        $\hat{\mathcal{O}}_7 = {\bf{\hat{S}}}_{N}\cdot {\bf{\hat{v}}}^{\perp}\mathds{1}_\chi$ & $A$\\ 
        $\hat{\mathcal{O}}_8 = {\bf{\hat{S}}}_{\chi}\cdot {\bf{\hat{v}}}^{\perp}\mathds{1}_N$  & $A$\\ 
        $\hat{\mathcal{O}}_9 = i{\bf{\hat{S}}}_\chi\cdot\left({\bf{\hat{S}}}_N\times\frac{{\bf{\hat{q}}}}{m_N}\right)$ & $B$\\ 
        $\hat{\mathcal{O}}_{10} = i{\bf{\hat{S}}}_N\cdot\frac{{\bf{\hat{q}}}}{m_N}\mathds{1}_\chi$   & $B$\\
        $\hat{\mathcal{O}}_{11} = i{\bf{\hat{S}}}_\chi\cdot\frac{{\bf{\hat{q}}}}{m_N}\mathds{1}_N$   & $B$\\
        $\hat{\mathcal{O}}_{12} = {\bf{\hat{S}}}_{\chi}\cdot \left({\bf{\hat{S}}}_{N} \times{\bf{\hat{v}}}^{\perp} \right)$  & $A$\\    
        $\hat{\mathcal{O}}_{13} =i \left({\bf{\hat{S}}}_{\chi}\cdot {\bf{\hat{v}}}^{\perp}\right)\left({\bf{\hat{S}}}_{N}\cdot \frac{{\bf{\hat{q}}}}{m_N}\right)$ & $B$\\ 
        $\hat{\mathcal{O}}_{14} = i\left({\bf{\hat{S}}}_{\chi}\cdot \frac{{\bf{\hat{q}}}}{m_N}\right)\left({\bf{\hat{S}}}_{N}\cdot {\bf{\hat{v}}}^{\perp}\right)$ & $B$\\          
        $\hat{\mathcal{O}}_{15} = -\left({\bf{\hat{S}}}_{\chi}\cdot \frac{{\bf{\hat{q}}}}{m_N}\right)\left[ \left({\bf{\hat{S}}}_{N}\times {\bf{\hat{v}}}^{\perp} \right) \cdot \frac{{\bf{\hat{q}}}}{m_N}\right] $  & $B$ \\ 
        $\hat{\mathcal{O}}_{17}=i \frac{{\bf{\hat{q}}}}{m_N} \cdot \mathbf{\mathcal{S}} \cdot {\bf{\hat{v}}}^{\perp} \mathds{1}_N$ & $B$\\          
$\hat{\mathcal{O}}_{18}=i \frac{{\bf{\hat{q}}}}{m_N} \cdot \mathbf{\mathcal{S}}  \cdot {\bf{\hat{S}}}_{N}$ & $B$\\                                                                     
    \bottomrule
      \bottomrule
      \boldmidrule
    \end{tabular}
    \end{ruledtabular}	
    \caption{Quantum mechanical operators defining the non-relativistic effective theory of DM-nucleon interactions~\cite{Fan:2010gt,Fitzpatrick:2012ix}.~The notation is the one introduced in Sec.~\ref{sec:theory}.~The operators $\hat{\mathcal{O}}_{1}$ and $\hat{\mathcal{O}}_{4}$ correspond to canonical spin-independent and spin-dependent interactions, respectively.~The operators $\hat{\mathcal{O}}_{17}$ and $\hat{\mathcal{O}}_{18}$ only arise for spin 1 WIMPs, and $\mathbf{\mathcal{S}}$ is a symmetric combination of spin 1 WIMP polarisation vectors~\cite{Dent:2015zpa}.~Operator $\hat{\mathcal{O}}_{2}$ is quadratic in ${\bf{\hat{v}}}^{\perp}$ and $\hat{\mathcal{O}}_{16}$ is a linear combination of $\hat{\mathcal{O}}_{12}$ and $\hat{\mathcal{O}}_{15}$ and are therefore not considered here~\cite{Anand:2013yka}.~Finally, the right column in the table shows the operator spectral type:~$A$ corresponds to featureless spectra and $B$ to bumpy spectra.}
    \label{tab:operators}
\end{table}

DM-nucleon interactions can therefore be described by the Lagrangian
\begin{equation}
\label{eq:LagrCoeffOp}
 \mathcal{L}_\text{int} = \sum_{N=n,p} \sum_i c_i^{(N)} \hat{\mathcal{O}}^{(N)}_i \,,
\end{equation}
where $i$ labels the interaction type, $N=p$ ($N=n$) denotes coupling to protons (neutrons), and the coupling constants $c_i^{(N)}$ have dimension mass to the power of $-2$.~In order to simplify the notation, from here onwards we will omit the nucleon index in the equations.~A detailed description of how to compute cross-sections for DM-nucleus scattering from Eq.~(\ref{eq:LagrCoeffOp}) can be found in \Refs~\cite{Fitzpatrick:2012ix,Anand:2013yka,Catena:2015uha}.~Limits on the coupling constants in~Eq.~(\ref{eq:LagrCoeffOp}) have been derived in, e.g.,~\cite{DelNobile:2013sia,Catena:2014uqa,Gresham:2014vja,Catena:2015uua,Catena:2015iea,Catena:2016kro,Aprile:2017aas}, and the prospects for DM particle detection in this framework have been studied in~\cite{Catena:2014hla,Catena:2014epa,Catena:2015vpa,Kavanagh:2016pyr,Catena:2016sfr,Catena:2016tlv,Catena:2017wzu}.

\begin{table}[t!]
    \centering
    \begin{ruledtabular}	
\begin{tabular}{lccc}
 \toprule
  \boldmidrule
  Spin 0 DM & Coeff. & Scalar med. & Vector med. \\ \addlinespace[.4em]
   & $c_1$ & $\frac{h_1^N g_1}{M^2_\Phi}$ & $-2 \frac{h_3^N g_4}  { M_G^2}$ \\ \addlinespace[.4em]
   & $c_7$ & & $4 \frac{h_4^N g_4}  { M_G^2}$   \\ \addlinespace[.4em]
   & $c_{10}$ & $\frac{h_2^N g_1}{M^2_\Phi}$ & \\
   \bottomrule
    \toprule
    \boldmidrule
    spin 1/2 DM & Coeff. & Scalar med. & Vector med. \\ \addlinespace[.4em]
    &$c_1$ & $\frac{h_1^N \lambda_1}{M^2_\Phi}$ & $-\frac{h_3^N \lambda_3}{M_G^2}$\\ \addlinespace[.4em]
    &$c_4$ & & $4\frac{h_4^N \lambda_4}{M_G^2}$\\ \addlinespace[.4em]
    &$c_6$ & $\frac{h_2^N \lambda_2}{M_\Phi^2}\frac{m_N}{m_\chi}$ & \\ \addlinespace[.4em]
    &$c_7$ & & $2\frac{h_4^N \lambda_3}{M_G^2}$\\ \addlinespace[.4em]
    &$c_8$ & & $-2\frac{h_3^N \lambda_4}{M_G^2}$\\ \addlinespace[.4em]
    &$c_9$ & & $-2\frac{h_4^N \lambda_3}{M_G^2}\frac{m_N}{m_\chi} - 2\frac{h_3^N \lambda_4}{M_G^2} $  \\ \addlinespace[.4em]   
    &$c_{10}$ & $\frac{h_2^N \lambda_1}{M^2_\Phi}$ & \\ \addlinespace[.4em]   
    &$c_{11}$ & $-\frac{h_1^N \lambda_2}{M_\Phi^2}\frac{m_N}{m_\chi}$ & \\
   \bottomrule
    \toprule
 \boldmidrule   
 Spin 1 DM & Coeff. & Scalar med. & Vector med. \\ \addlinespace[.4em]
&   $c_1$ & $\frac{b_1 h_1^N}{M^2_\Phi}$ & $-2 \frac{h_3^N b_5}  { M_G^2}$ \\ \addlinespace[.4em]
 &  $c_4$ & & \hspace{-0.15 cm}$-4\frac{h_4^N \Re(b_7)}{M_G^2} +\frac{q^2}{m_X m_N} \frac{h_3^N \Im(b_6)}{M_G^2}$ \\ \addlinespace[.4em] 
 &  $c_5$ &  & $-\frac{m_N}{m_X}\frac{h_3^N \Im(b_6)}{M_G^2}$ \\ \addlinespace[.4em]
 &  $c_6$ &  &  $-\frac{m_N}{m_X}\frac{h_3^N \Im(b_6)}{M_G^2}$ \\ \addlinespace[.4em]
 & $c_7$ & & $4\frac{h_4^N b_5}  { M_G^2}$   \\ \addlinespace[.4em]
 & $c_8$ &  & $2\frac{h_3^N \Re(b_7)}{M_G^2}$ \\ \addlinespace[.4em]
 & $c_9$ &  & $-2\frac{m_N}{m_X}\frac{h_4^N \Im(b_6)}{M_G^2} + 2\frac{h_3^N \Re(b_7)}{M_G^2}$ \\ \addlinespace[.4em]
 & $c_{10}$ & $\frac{b_1 h_2^N}{M_\Phi^2}$ & \\ \addlinespace[.4em]
 & $c_{11}$ & & $-\frac{m_N}{m_X}\frac{h_3^N \Im(b_7)}{M_G^2}$ \\ 
 & $c_{14}$ & & $2 \frac{m_N}{m_X}\frac{h_4^N \Im(b_7)}{M_G^2}$ \\ 
     \bottomrule
     \bottomrule
    \boldmidrule
    \end{tabular}    
    \end{ruledtabular}	
    \caption{Relation between the coupling constants of non-relativistic operators from \reftab{tab:operators} (in the proton/neutron basis) and simplified models in this study (see \Ref~\cite{Dent:2015zpa} for a full list).~For simplicity, in the second column we omit the index $N$.~In the case of spin 1 DM, we do not consider operators depending on the symmetric combination of polarisation vectors $\mathcal{S}$.}
    \label{tab:coeffs}
\end{table}

\subsection{Connecting non-relativistic operators and simplified models}
The non-relativistic operators in \reftab{tab:operators} can arise from the non-relativistic reduction of the simplified models discussed above~\cite{Dent:2015zpa,DelNobile:2013sia}.~One can therefore directly translate the parameters of a simplified model into the coefficients given in \refeq{eq:LagrCoeffOp}, as has been shown in \Ref~\cite{Dent:2015zpa} in detail.~For instance, if we integrate out the heavy vector mediator $G^\mu$ in Eq.~(\ref{eq:LagrGchi}), we obtain $c^{(N)}_1 = -h_3^N \lambda_3/M_G^2$ and $c^{(N)}_4 = 4 h_4^N \lambda_4/M_G^2$.~The couplings to nucleons $h_i^N$ are related to the quark level couplings by nucleon form factors~\cite{Dent:2015zpa,Agrawal:2010fh,Dienes:2013xya}:
\begin{subequations}
 \begin{align}
 &h_1^n = 11.93 \, h_1	& &h_1^p = 12.31 \, h_1 \\
 &h_2^n = -0.07 \, h_2	& &h_2^p = -0.28 \, h_2  \\
 &h_3^{n,p} = 3 \, h_3 &
 &h_4^{n,p} = 0.33 \, h_4\,.
\end{align}
\end{subequations}
The coefficients $c_i^{(N)}$ for all other relevant cases are listed in \reftab{tab:coeffs}\footnote{Some of the coefficients in Tab.~\ref{tab:coeffs} differ from those in the published version of Ref.~\cite{Dent:2015zpa}.~A revised version of Ref.~\cite{Dent:2015zpa} is currently in preparation.~The coefficients in the revised version will agree with Tab.~\ref{tab:coeffs} and Ref.~\cite{DelNobile:2013sia}.}.~Sub-leading operators that might arise in the non-relativistic reduction are not reported in Tab.~\ref{tab:coeffs}.~Furthermore, when multiple operators are associated to a simplified model, we only consider the leading one:~$\hat{\mathcal{O}}_7$ in the case $h_4\neq0$ and $\lambda_3\neq0$, and $\hat{\mathcal{O}}_8$ in the case $h_3\neq0$ and $\lambda_4\neq0$.~As already mentioned, here we assume that all quark level couplings are universal\footnote{This means that also for scalar mediators all couplings will have the same value and are not proportional to the Standard Model Yukawa couplings, which is another scenario often studied in literature.} and leave the study of non-universal coupling corrections for future investigations.~Note, however, that due to the different nucleon form factors, couplings to neutrons and protons can differ\footnote{This is not surprising, since isospin is broken at the scale of nucleons, as we know from the different masses of protons and neutrons.}.~The matching procedure that we have outlined above does not account for two potentially important phenomena:~1) Operator evolution from the mediator mass scale down to the nuclear recoil energy scale~\cite{Crivellin:2014qxa,DEramo:2014nmf}; 2) Momentum-dependent chiral effective field theory corrections to the $c_i^{(N)}$ coefficients induced by inter-nucleon interactions mediated by meson exchange~\cite{Bishara:2016hek,Hoferichter:2015ipa}.~Both effects have only been studied for spin $\le 1/2$ DM, and extending these studies to spin 1 DM goes beyond the scope of the present work.~Nevertheless, we will briefly comment on how operator evolution can affect our results in Sec.~\ref{sec:discussion}.

\begin{figure}[t]
\label{fig:spectra}
 \begin{center}
  \includegraphics[width=\linewidth]{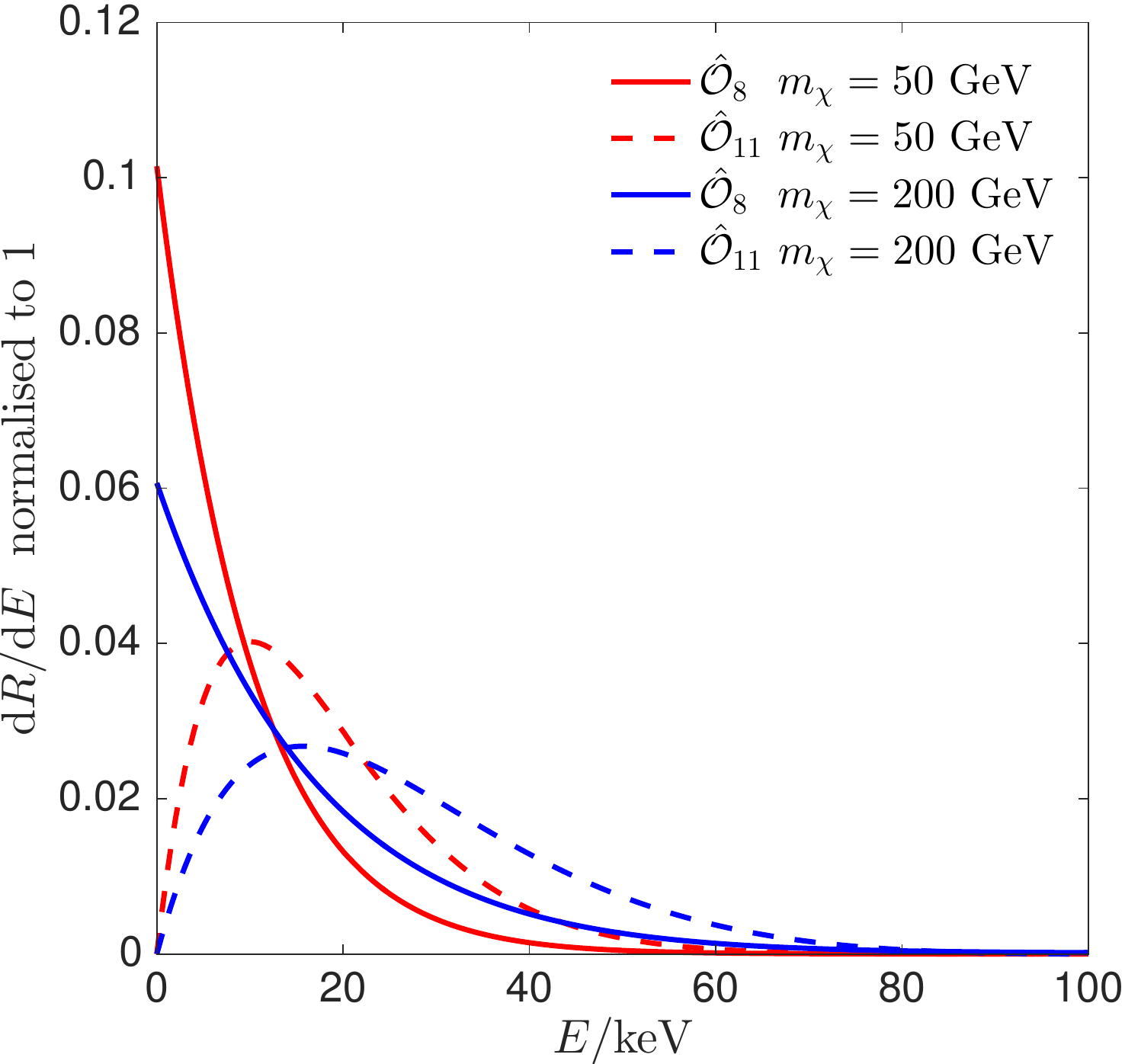}
 \end{center}
\caption{Nuclear recoil energy spectra for selected operators and masses.~The non-relativistic operator $\hat{\mathcal{O}}_8$ generates a spectrum of type A, whereas the operator $\hat{\mathcal{O}}_{11}$ generates a spectrum of type B.\label{fig:spectra}}
\end{figure}

\section{DM detection at XENONNT}
\label{sec:detection}
In this section we describe the main features of a DM signal at XENONnT, and show how these can be simulated numerically.~In the next section we will explore the consequences of a DM signal at XENONnT for the mono-jet searches that will be performed at the LHC during Run 3.
\subsection{Expected S1 signal}
\label{sec:S1signal}
DM detection at XENONnT would occur through the observation of scintillation photons emitted by the deexcitation of Xe$_2^*$ molecules produced by the passage of DM particles in liquid Xenon.~The observation of scintillation photons is performed via conversion into photoelectrons in a time projection chamber (TPC).~The number of photoelectrons produced per DM interaction by the prompt deexcitation of Xe$_2^*$ molecules is denoted by S1.~A secondary scintillation (S2) signal produced by electrons generated in the DM scattering and drifted to the top of the detector by an electric field is used for background discrimination but will be neglected here for simplicity.~This simplification is motivated by the fact that the S1 and S2 signals are anti-correlated~\cite{Aprile:2015uzo}.~The expected rate of DM-induced S1 events per unit detector mass is given by~\cite{Aprile:2011hx}
\begin{eqnarray}
\frac{{\rm d}R}{{\rm d}{\rm S1}} &=& \zeta(S1) \sum_{n=1}^{\infty}\mathscr{G}({\rm S1}| n,\sqrt{n}\hat{\sigma}) 
 \int_0^\infty {\rm d}E\, \frac{{\rm d}R}{{\rm d}E} 
 \mathscr{P}(n|\nu(E)) \,, \nonumber\\
\label{eq:rateS1}
\end{eqnarray}
where $\mathscr{G}$ is a Gaussian distribution of mean $n$ and variance $\sqrt{n}\hat{\sigma}$,  $\mathscr{P}$ is a Poisson distribution of mean $\nu(E)$, and ${\rm d}R/{\rm d}E$ is the rate of nuclear recoil events per unit detector mass
\begin{equation}
\frac{{\rm d}R}{{\rm d}E} = \sum_T  \frac{\xi_T\rho_\chi}{m_T m_\chi} \int_{|\mathbf{v}|<v_{\rm esc}} {\rm d}^3\mathbf{v}\, |\mathbf{v}|f(\mathbf{v}+\mathbf{v}_{\oplus}) \frac{{\rm d}\sigma_T(E,|\mathbf{v}|)}{{\rm d}E}\,.
\label{eq:rateE}
\end{equation}
Here $\nu(E)$ is the number of photoelectrons expected when a nuclear recoil energy $E$ is deposited in the detector (given in Fig.~13 of~\cite{Aprile:2015uzo}), $n$ (S1) is the number of actually produced (observed) photoelectrons, $\hat{\sigma}=0.4$ is the single-photoelectron resolution of the XENONnT photomultipliers, and $\zeta(S1)\simeq 0.4$ is the experimental acceptance~\cite{Aprile:2015uzo}.~In Eq.~(\ref{eq:rateE}), $f$ is the DM velocity distribution in the galactic rest frame boosted to the detector rest frame, $\mathbf{v}_\oplus$ is the Earth's velocity in the galactic rest frame, and $\rho_\chi$ is the local DM density.~We set $\rho_\chi$ to 0.3~GeV~cm$^{-3}$ and assume a Gaussian distribution truncated at the galactic escape velocity $v_{\rm esc}=533$~km~s$^{-1}$ for $f$, as in the so-called Standard Halo Model~\cite{Freese:2012xd}.~Different choices might be considered, however~\cite{Catena:2009mf,Catena:2011kv}.~The differential cross-section ${\rm d}\sigma_T(E,|\mathbf{v}|)/{\rm d}E$ in Eq.~(\ref{eq:rateE}) is calculated for each simplified model considered in this work using nuclear response functions computed in~\cite{Anand:2013yka}.~The sum in Eq.~(\ref{eq:rateE}) is performed over the seven most abundant Xenon isotopes.~Isotope masses and mass fractions are denoted by $m_T$ and $\xi_T$, respectively.~The number of signal events is obtained by integrating Eq.~(\ref{eq:rateS1}) from S1=3 to S1=70, and multiplying the result by the experimental exposure $\varepsilon$.~Here we assume $\varepsilon=20$~ton$\times$year.~An analysis extending to S1=180 has recently been published by the XENON100 collaboration~\cite{Aprile:2017aas} and will also be repeated with the actual XENONnT data.~Finally, S1 background events at XENONnT are modelled according to Fig.~14 in~\cite{Aprile:2015uzo}.~Fig.~\ref{fig:spectra} shows the ${\rm d}R/{\rm d}E$ spectra for selected DM-nucleon interaction operators in Tab.~\ref{tab:operators}.~As already anticipated in Sec.~\ref{sec:introduction}, energy spectra divide into two categories:~spectra of Type $A$ with a maximum at $E=0$ ($q=0$), and spectra of type $B$ with a maximum at $E\neq0$ ($q\neq 0$). 
 
\begin{table}[t]
\centering
    \begin{ruledtabular}
    \begin{tabular}{lcccc}
    \toprule
\boldmidrule	
 Scalar DM & Op. & $g_q$ & $g_\text{DM}$ & $M_\text{eff}$ [GeV]\\
&1	&$h_1$	&$g_1$	&14564.484	\\
&1	&$h_3$	&$g_4$	&10260.217	\\
&7	&$h_4$	&$g_4$	&4.509	\\
&10	&$h_2$	&$g_1$	&10.706	\\
 \toprule
\boldmidrule
 Fermionic DM & Op. & $g_q$ & $g_\text{DM}$ & $M_\text{eff}$ [GeV]\\
&1	&$h_1$	&$\lambda_1$	&14564.484	\\
&1	&$h_3$	&$\lambda_3$	&7255.068	\\
&4	&$h_4$	&$\lambda_4$	&147.354	\\
&6	&$h_2$	&$\lambda_2$	&0.286	\\
&7	&$h_4$	&$\lambda_3$	&3.188	\\
&8	&$h_3$	&$\lambda_4$	&225.159	\\
&10	&$h_2$	&$\lambda_1$	&10.706	\\
&11	&$h_1$	&$\lambda_2$	&351.589	\\
\toprule
\boldmidrule
Vector DM & Op. & $g_q$ & $g_\text{DM}$ & $M_\text{eff}$ [GeV]\\
&1	&$h_1$	&$b_1$	&14564.484	\\
&1	&$h_3$	&$b_5$	&10260.216	\\
&4	&$h_4$	&$\Re$($b_7$)	&188.302	\\
&5	&$h_3$	&$\Im$($b_6$)	& 6.946 \\
&7	&$h_4$	&$b_5$	&4.509	\\
&8	&$h_3$	&$\Re$($b_7$)	&287.728	\\
&9	&$h_4$	&$\Im$($b_6$)	&3.674	\\
&10	&$h_2$	&$b_1$	&10.706	\\
&11	&$h_3$	&$\Im$($b_7$)	&223.794	\\
&14	&$h_4$	&$\Im$($b_7$)	&0.201	\\
  \boldmidrule
\bottomrule
 \end{tabular}    	
 \end{ruledtabular}	
\caption{Benchmark points producing 150 signal events in an idealised version of XENONnT (see text at the beginning of Sec.~\ref{sec:forecasts}) for $m_\chi = 50\gev$.}
\label{tab:benchmarks}
\end{table}

\subsection{Simulated S1 signal}
Using Eq.~(\ref{eq:rateS1}) and the background model in~\cite{Aprile:2015uzo}, we can now simulate the detection of S1 events at XENONnT.~We simulate S1 events at XENONnT using standard Monte Carlo methods, e.g.~\cite{Cowan:2010js}.~If $\mu_{\rm S}$ is the expected number of signal events, we sample the actual number of observed S1 events, $N_{\rm exp}$, from a Poisson distribution of mean $\mu_{\rm tot}=\mu_{\rm S}+\mu_B$, where $\mu_B\simeq 41$ is the expected number of background events at XENONnT in the (3,70) signal region~\cite{Aprile:2015uzo}.~S1 values for the $N_{\rm exp}$ observed events are sampled from the following probability density function (PDF)
\begin{equation}
f_{\rm tot}({\rm S}1) = \mathcal{N}\left(\frac{{\rm d}R}{{\rm d}{\rm S}1}\bigg|_{\rm signal} + \frac{{\rm d}R}{{\rm d}{\rm S}1}\bigg|_{\rm background} \right)\,,
\label{eq:ftot}
\end{equation}
where the signal contribution to $f_{\rm tot}$ is computed using Eq.~(\ref{eq:rateS1}) and the background contribution to $f_{\rm tot}$ is extracted from Fig.~14 in~\cite{Aprile:2015uzo}.~The normalisation constant $\mathcal{N}$ is defined by 
\begin{equation}
\int_{{\rm S}1=3}^{{\rm S}1=70} {\rm d}{\rm S}1\,f_{\rm tot}({\rm S}1)=1\,.
\end{equation}
We simulate S1 events from 22 distinct simplified models, each characterised by a pair of coupling constants $g_q$ and $g_{\rm DM}$ (defined below) and by the quantum numbers, non-relativistic limit and Lagrangians given in Tab.~\ref{tab:benchmarks} and in \app~\ref{app:Lagrangians}.~Here, $g_q$ refers to the mediator-quark-antiquark vertex and $g_{\rm DM}$ is associated with the mediator-DM-DM vertex.~For example, in the case of scalar DM coupled to quarks via scalar exchange, $g_q=h_1$ and $g_{\rm DM}=g_1$.~For each simplified model that we consider, Tab.~\ref{tab:benchmarks} also shows benchmark values for the effective mass, $M_{\rm eff}$, 
\begin{equation}
M_\text{eff} \equiv M_{\rm med} \sqrt{\frac{(0.1)^2}{g_q g_\text{DM}}}\,,
\label{eq:Meff}
\end{equation}
obtained as explained at the beginning of Sec.~\ref{sec:forecasts}.~In Eq.~(\ref{eq:Meff}), $M_{\rm med}$ generically denotes the mediator mass in the given simplified model.~We normalise $M_{\rm eff}$ in such a way that $M_{\rm eff}=M_{\rm med}$ for $g_q=g_{\rm DM}=0.1$, and choose $g_q=g_{\rm DM}=0.1$ as typical values for weak-scale couplings.~From Tab.~\ref{tab:coeffs}, we see that direct detection is sensitive to the effective mass $M_{\rm eff}$, and not to $g_q$, $g_{\rm DM}$ and $M_{\rm med}$ separately.~Below, a model characterised by $g_q\neq0$ and $g_{\rm DM}\neq0$, and where DM predominantly interacts with nucleons via the operator $\hat{\mathcal{O}}_i$, will be denoted by $\hat{\mathcal{O}}_i(g_q,g_{\rm DM})$.

\begin{figure}[t]
 \begin{center}
  \includegraphics[width=\linewidth]{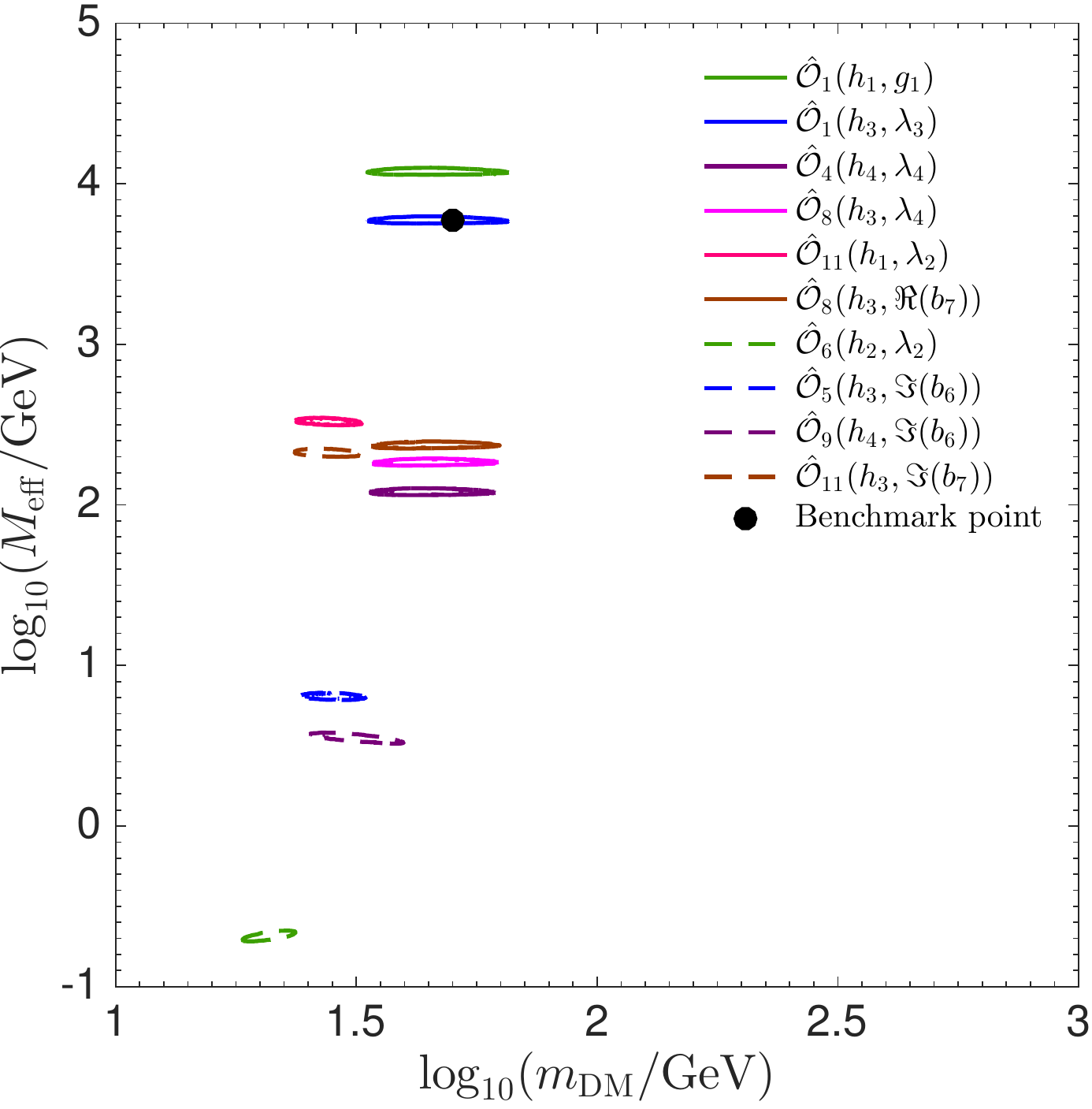}
 \end{center}
\caption{2D 95\% confidence intervals in the $m_{\rm DM} - M_{\rm eff}$ plane obtained by fitting selected models from Tab.~\ref{tab:benchmarks} to data simulated from model $\hat{\mathcal{O}}(h_3,\lambda_3)$.~In the simulation, we set $m_{\rm DM}=50$~GeV and $\mu_{\rm S}\simeq 100$.~As one can appreciate from this figure, the error on $M_{\rm eff}$ is small compared to the various astrophysical and collider uncertainties entering our analysis.~\label{fig:DeltaNS}}
\end{figure}

\subsection{Parameter inference with XENONnT}
\label{sec:modsel}
We now focus on the determination of $M_{\rm eff}\propto \mu_{\rm S}^{-1/4}$ at XENONnT.~As we will see in Sec.~\ref{sec:LHC}, knowing $M_{\rm eff}$ it is possible to predict if a given simplified model is expected to produce an observable mono-jet signal at the LHC Run 3, and if so, where in the $M_{\rm med} - \sigma$ plane.~Specifically, here we show that in the analysis of $\mathcal{O}(100)$ S1 events at XENONnT, the true value of $\mu_{\rm S}\propto M^{-4}_{\rm eff}$ is always reconstructed within a relative error of about $\sim 30$\% or less, independently of the simplified model underlying the data and of the model assumed in the fit.~From the reconstructed value of $\mu_{\rm S}$ and for each model in Tab.~\ref{tab:benchmarks}, one can then obtain $M_{\rm eff}$ within the same accuracy.~It is a typical parameter inference problem, which we address as follows.~From a given model in Tab.~\ref{tab:benchmarks}, we simulate a sample of $N_{\rm exp}$ S1 events at XENONnT, as described in the previous section.~We then fit all models in Tab.~\ref{tab:benchmarks} to the $N_{\rm exp}+1$ data points, maximising the likelihood  function
\begin{align}
\mathscr{L}(\boldsymbol{d}\,|\,\boldsymbol{\Theta},\mathcal{H})&= \frac{\mu_{\rm tot}(\boldsymbol{\Theta},\mathcal{H})^{N_{\rm exp}(\boldsymbol{d})}}{N_{\rm exp}(\boldsymbol{d})!} \,e^{-\left[\mu_{\rm tot}(\boldsymbol{\Theta},\mathcal{H})\right]}\nonumber\\
&\times\prod_{i=1}^{N_{\rm exp}(\boldsymbol{d})} f_{\rm tot}(x_i(\boldsymbol{d}),\mathcal{H})\,,
\label{eq:like}
\end{align}
where $\mathbf{d}$ is an array of data including the number of observed events, $N_{\rm exp}$, and the number of observed photoelectrons in a given event, $x_i$, $i=1,\dots,N_{\rm exp}$.~In Eq.~(\ref{eq:like}), $\boldsymbol{\Theta}=(m_{\rm DM},M_{\rm eff})$, and $\mathcal{H}$ is the hypothesis made in the fit regarding the DM particle spin and non-relativistic interactions\footnote{For simplicity, in Eq.~(\ref{eq:ftot}) we have omitted the $\mathcal{H}$ dependence of the function $f_{\rm tot}$.}.~Results are presented in terms of confidence intervals in the $m_{\rm DM} - M_{\rm eff}$ plane.~For example, Fig.~\ref{fig:DeltaNS} shows the 2D 95\% confidence intervals in the $m_{\rm DM} - M_{\rm eff}$ plane that we obtain by fitting selected simplified models from Tab.~\ref{tab:benchmarks} to data $\boldsymbol{d}$ generated from model $\hat{\mathcal{O}}_{1}(h_3,\lambda_3)$.~In this specific calculation, we assume $\mu_{\rm S}\simeq 100$, unlike Tab.~\ref{tab:benchmarks}.~The fit is performed using {\sffamily multinest}~\cite{Feroz:2008xx}, and confidence intervals are computed with a modified version of {\sffamily SuperBayes}, e.g.~\cite{2011JHEP...06..042F}.~As one can appreciate from this figure, the error on $M_{\rm eff}$ is small compared to the various uncertainties entering our analysis, including those affecting many astrophysical and collider inputs.~In this example, data were generated from model $\hat{\mathcal{O}}_{1}(h_3,\lambda_3)$ in Tab.~\ref{tab:benchmarks}.~However, the same conclusion applies to fits to data generated from different simplified models (or to different realisations of the same data).~Therefore, if XENONnT will observe $\mathcal{O}(100)$ signal events, the true value of $\mu_{\rm S}$, and associated values of $M_{\rm eff}$, will likely be reconstructed with an error that is negligible for our purposes.~We will use this result in Sec.~\ref{sec:LHC}.

\begin{figure*}[t]
\begin{center}
\begin{minipage}[t]{0.49\linewidth}
\centering
\includegraphics[width=\textwidth]{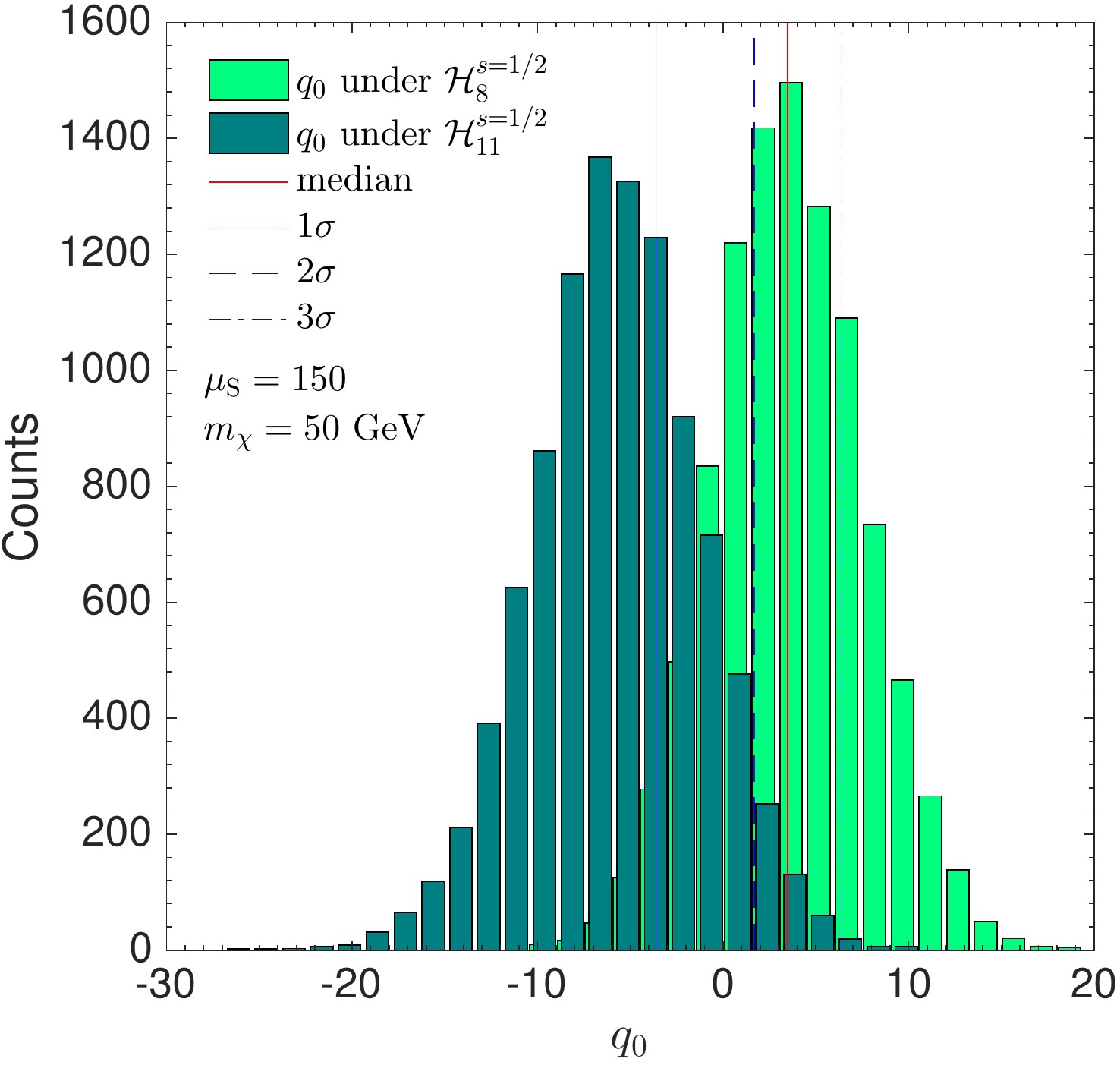}
\end{minipage}
\begin{minipage}[t]{0.49\linewidth}
\centering
\includegraphics[width=\textwidth]{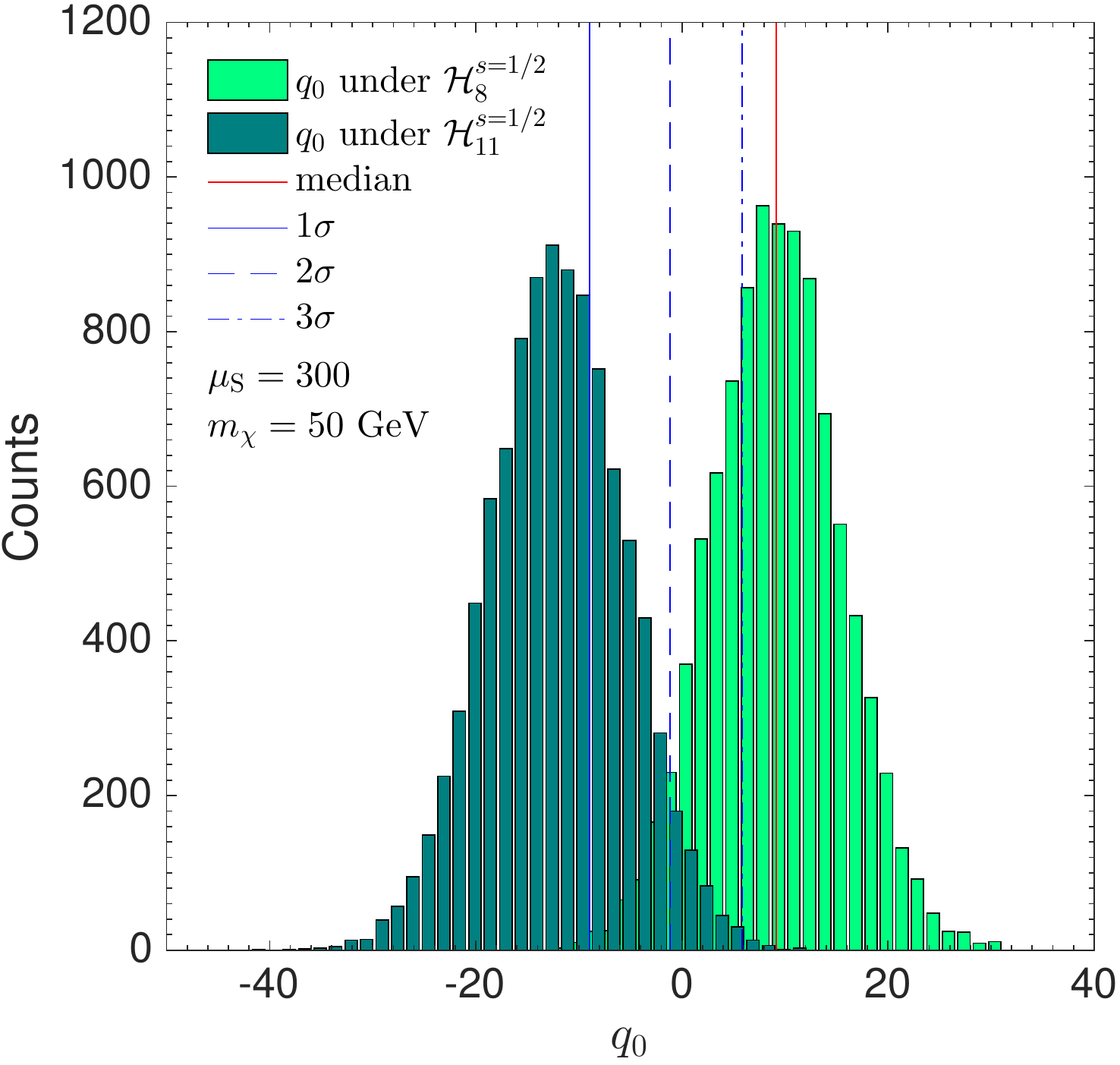}
\end{minipage}
\begin{minipage}[t]{0.49\linewidth}
\centering
\includegraphics[width=\textwidth]{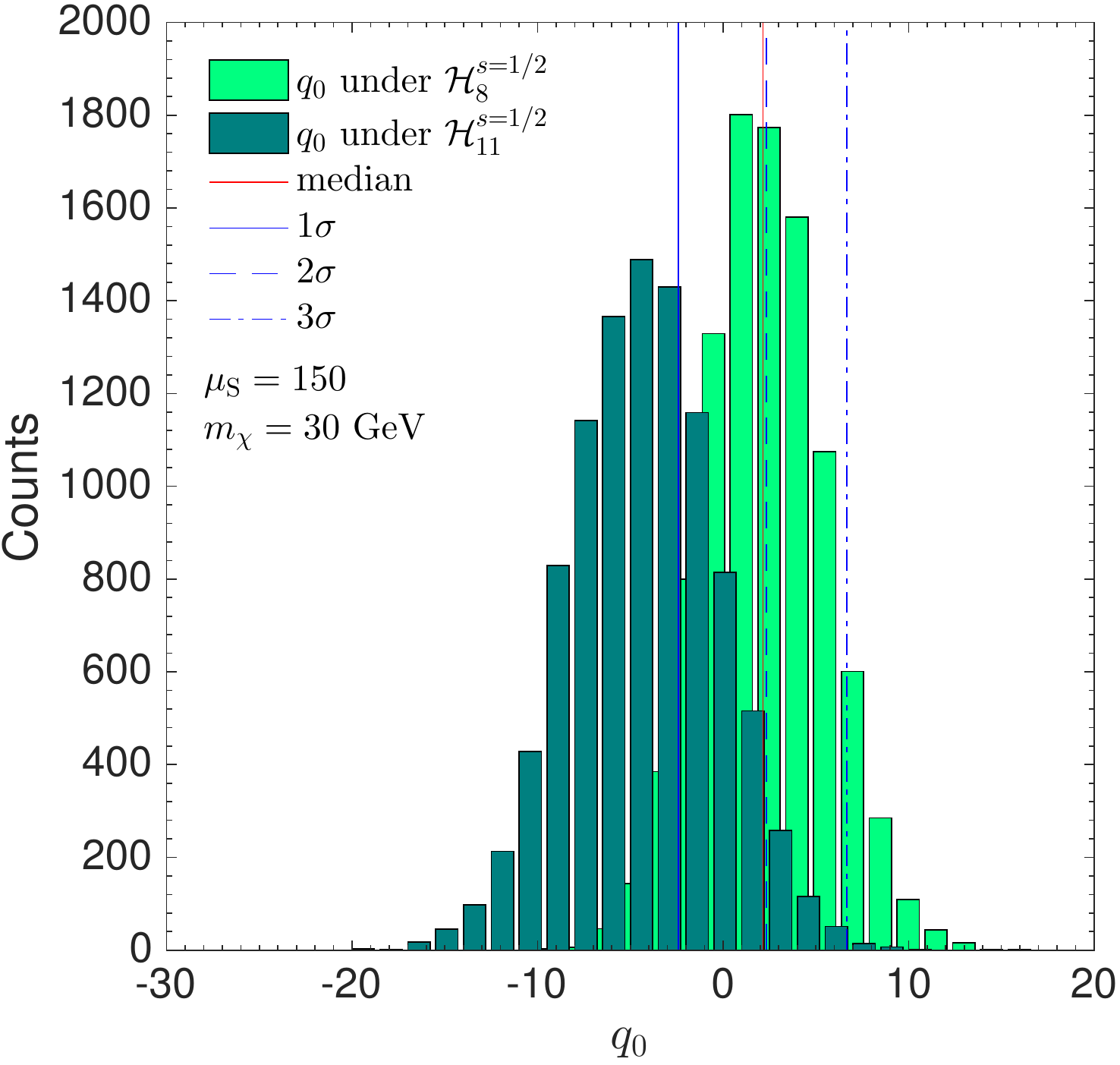}
\end{minipage}
\begin{minipage}[t]{0.49\linewidth}
\centering
\includegraphics[width=\textwidth]{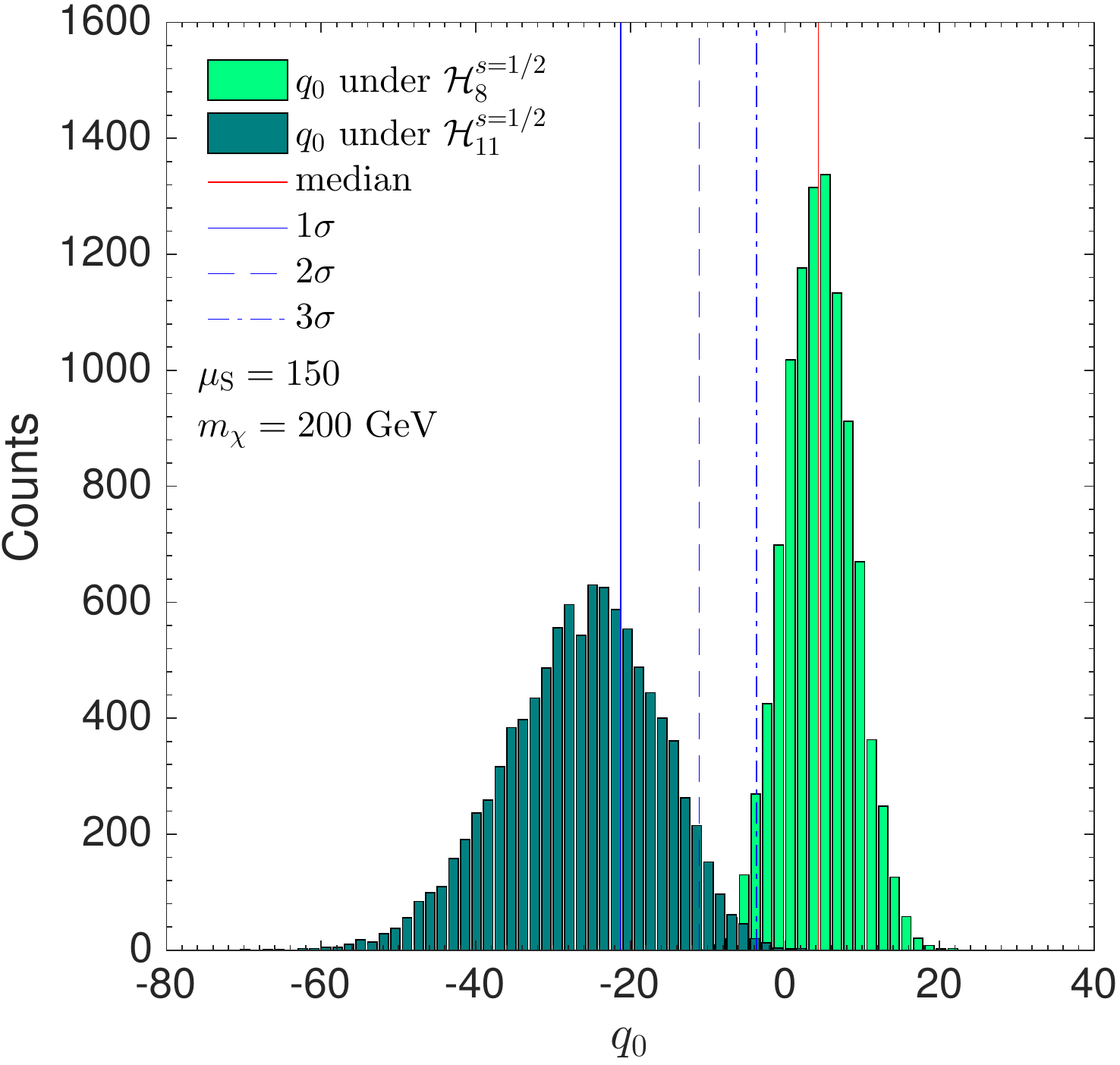}
\end{minipage}
\end{center}
\caption{PDFs $f(q_0|\boldsymbol{d}_{\mathcal{H}_0})$ and $f(q_0|\boldsymbol{d}_{\mathcal{H}_a})$ obtained for the hypotheses $\mathcal{H}_0=\mathcal{H}_{8}^{s=1/2}$ (spin 1/2 DM interacting through $\hat{\mathcal{O}}_8$) and $\mathcal{H}_a=\mathcal{H}_{11}^{s=1/2}$ (spin 1/2 DM interacting through $\hat{\mathcal{O}}_{11}$) and different choices of $\mu_{\rm S}$ and of the DM particle mass $m_{\rm DM}$.~For each choice of input parameters, the four panels also show the median of the $f(q_0|\boldsymbol{d}_{\mathcal{H}_a})$ PDF (red curve), and the $1\sigma$, $2\sigma$ and $3\sigma$ thresholds $q_{n\sigma}$, $n=1,2,3$ (blue curves), derived from $f(q_0|\boldsymbol{d}_{\mathcal{H}_0})$.\label{fig:selection}}  
\end{figure*}

\begin{figure}[t]
 \begin{center}
  \includegraphics[width=0.8\linewidth]{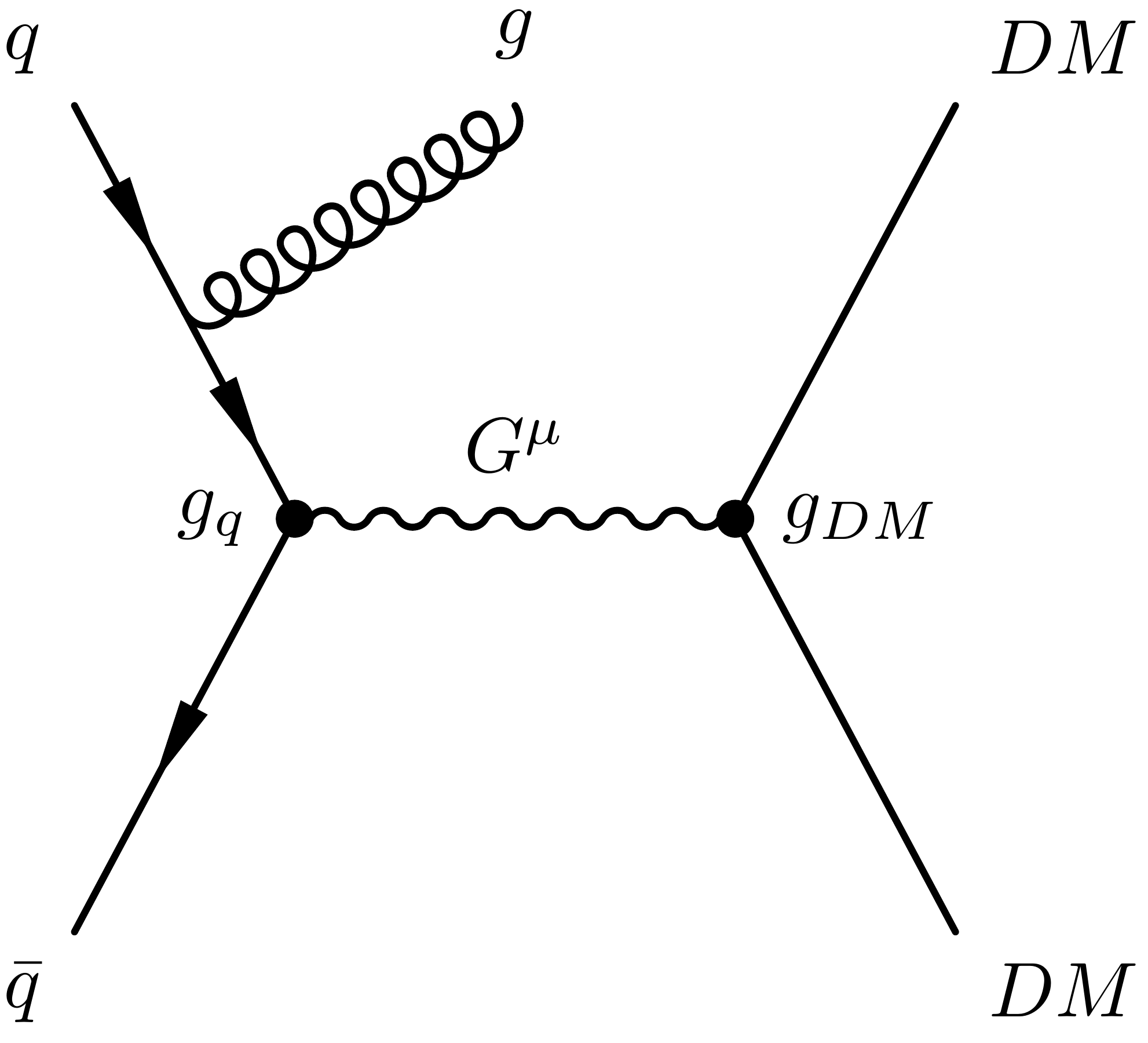}
 \end{center}
\caption{Example of a diagram where a gluon is emitted from an initial state quark, leading to mono-jet after hadronization and showering of the gluon.\label{fig:jet}}
\end{figure}

\subsection{DM model selection with XENONnT}
\label{sec:modsel}
In this section we show that $\mathcal{O}(100)$ events at XENONnT are enough to statistically discriminate featureless spectra of type $A$ from bumpy spectra of type $B$.~As we will see in Sec.~\ref{sec:LHC}, this result will allow us to significantly narrow the regions in the $M_{\rm med} - \sigma$ plane where a mono-jet signal is expected.~It is a non-nested model selection problem, that we address as follows.~Let us denote by $\mathcal{H}_0$ the hypothesis, i.e. the simplified model, that we would like to reject, and by $\mathcal{H}_a$ the alternative hypothesis.~We base model selection upon the following test statistic 
\begin{align}
q_0=-2\ln\left[\frac{\mathscr{L}(\boldsymbol{d}\,|\,\widehat{\boldsymbol{\Theta}}_0,\mathcal{H}_0)}{\mathscr{L}(\boldsymbol{d}\,|\,\widehat{\boldsymbol{\Theta}}_a,\mathcal{H}_a)}\right] \,,
\label{eq:q0}
\end{align}
where $\mathscr{L}$ is the likelihood  function of the simulated data $\mathbf{d}$, and $\widehat{\boldsymbol{\Theta}}_0$ ($\widehat{\boldsymbol{\Theta}}_a$) is the value of $\boldsymbol{\Theta}_0$ ($\boldsymbol{\Theta}_a$) that maximises the likelihood  $\mathscr{L}$ when fitting the data $\boldsymbol{d}$ under the hypothesis $\mathcal{H}_0$ ($\mathcal{H}_a$).~For $\mathscr{L}$, we assume Eq.~(\ref{eq:like}).~Given the test statistic  in Eq.~(\ref{eq:q0}), the statistical significance with which $\mathcal{H}_0$ can be rejected is computed as follows.~For each value of $\mu_{\rm S}$ that we consider, we simulate 10000 pseudo-experiments under the hypothesis $\mathcal{H}_a$.~We then construct the PDF of $q_0$ under $\mathcal{H}_a$, $f(q_0|\boldsymbol{d}_{\mathcal{H}_a})$, and calculate the associated median, $q_{\rm med}$.~$q_{\rm med}$ represents the ``typical'' value of $q_0$ when DM interacts according to $\mathcal{H}_a$.~Subsequently, we simulate 10000 pseudo-experiments under the hypothesis $\mathcal{H}_0$.~From these simulations we obtain the PDF of $q_0$ under $\mathcal{H}_0$, $f(q_0|\boldsymbol{d}_{\mathcal{H}_0})$, and solve for $q_{n\sigma}$ the integral equation
\begin{align}
1-\alpha_n=\int_{q_{n\sigma}}^\infty{\rm d}q_0 \,f(q_0|\boldsymbol{d}_{\mathcal{H}_0})\,.
\label{eq:p}
\end{align}
where $\alpha_1=0.6827$, $\alpha_2=0.9545$, and $\alpha_3=0.9973$.~By comparing $q_{\rm med}$ with $q_{n\sigma}$, we obtain a measure of the statistical significance with which the $\mathcal{H}_0$ hypothesis can be rejected.~For example, for $q_{\rm med}=q_{2\sigma}$ data imply a $2\sigma$ rejection of $\mathcal{H}_0$.

We now apply the method illustrated above to the case in which the hypothesis $\mathcal{H}_0=\mathcal{H}_{8}^{s=1/2}$ assumes spin 1/2 DM interacting through $\hat{\mathcal{O}}_8$ and $\mathcal{H}_a=\mathcal{H}_{11}^{s=1/2}$ assumes spin 1/2 DM interacting through $\hat{\mathcal{O}}_{11}$.~Fig.~\ref{fig:selection} shows the PDFs $f(q_0|\boldsymbol{d}_{\mathcal{H}_0})$ and $f(q_0|\boldsymbol{d}_{\mathcal{H}_a})$ that we obtain in this case for different choices of $\mu_{\rm S}$ and of the DM particle mass $m_{\rm DM}$.~For each choice of input parameters, Fig.~\ref{fig:selection} also shows the median of the $f(q_0|\boldsymbol{d}_{\mathcal{H}_a})$ PDF, and the $1\sigma$, $2\sigma$ and $3\sigma$ thresholds $q_{n\sigma}$, $n=1,2,3$, derived from $f(q_0|\boldsymbol{d}_{\mathcal{H}_0})$.~For $m_{\rm DM}=30$~GeV, 150 signal events at XENONnT allows to discriminate spectra of type $A$ from spectra of type $B$ with a statistical significance of $\sim 2\sigma$.~For smaller masses the statistical significance decreases, since spectral differences are to a large extent below the experimental threshold.~For larger masses the statistical significance of the rejection increases, being larger than 3$\sigma$ for $m_{\rm DM}=200$~GeV.~The results illustrated in this specific example are general, and also apply to other pairs of spectra of type $A$ and $B$.~We will use this result in Sec.~\ref{sec:LHC}.

\section{Impact on LHC mono-jet searches}
\label{sec:LHC}
In this section, we will show that the hypothetical detection of $\mathcal{O}(100)$ events at XENONnT can only be explained within relatively narrow model dependent regions of the $M_{\rm med} - \sigma$ plane.~Comparing these regions to the LHC current limits and projected sensitivity, we will forecast the outcome of the LHC Run 3 based on the hypothetical XENONnT signal.~For the simplified model in Tab.~\ref{tab:benchmarks} to be detected at XENONnT, the mediator necessarily couples to Standard Model quarks.~Hence, the mediator can be produced in proton-proton collision at the LHC and decay to a pair of DM particles which will escape the detector. If an additional gluon or quark is emitted from the initial state (see, e.g., Fig.~\ref{fig:jet}), this will lead to a single jet plus missing transverse energy $\ETmiss$ in the detector~\cite{ATL-PHYS-PUB-2014-007,Aaboud:2016tnv,CMS-PAS-EXO-16-037,CMS-PAS-EXO-16-013,CMS-DP-2016-057,Chala:2015ama,Buchmueller:2014yoa}, the so-called ``mono-jet'' signature we will focus on in the following.

In the following, the LHC current limits on and projected sensitivity to $\sigma$ are presented in Sec.~\ref{sec:limits}, our mono-jet simulations are described in Sec.~\ref{sec:jetsim}, while our XENONnT-based forecasts for the LHC Run 3 are illustrated in Sec.~\ref{sec:forecasts}. 

\subsection{Current limits and projections}
\label{sec:limits}
For vector, axial-vector, scalar and pseudo-scalar mediators, and for selected values of the coupling constants, current limits on the mono-jet production cross-section $\sigma$ can be extracted from \Ref~\cite{CMS-PAS-EXO-16-037} for CMS and from \Ref~\cite{Aaboud:2016tnv} for ATLAS.~For different simplified models and parameter values, limits on $\sigma$ are not available.~However, since we only consider final states involving a hadronic jet and $\slashed{E}_T$, we can assume that upper limits on the fiducial cross section $\sigma_{\rm fid.} \equiv \sigma \times {\cal A}$ are approximately universal, i.e. model independent, where ${\cal A}$ is the detector acceptance after selection-cuts.\footnote{Here we derived LHC constraints on the fiducial cross section.~Considering differential cross sections in terms of the jet $p_T$ and comparing background and signal prediction bin by bin would make our constraints more robust, but we leave this second approach for future work.}~Comparing predictions for different simplified models, it is therefore convenient to focus on the $M_{\rm med} - (\sigma\times \mathscr{A})$ plane.~In this plane, all simplified models are subject to the same constraint, which can be obtained computing the upper limit on $\sigma\times \mathscr{A}$ at a given point in the parameter space of a reference simplified model.~For example, the 95\% CL  exclusion limit on $M_{\rm med}$ for an axial-vector mediator with couplings $\gDM = 1$ and $g_q = 0.25$ is $M_{\rm med} \gtrsim 1950 \gev$ for $m_\text{DM} \lesssim 100 \gev$~\cite{CMS-PAS-EXO-16-037} (here we focus on CMS limits, since these are more constraining).~For the integrated luminosity of $12.9 \fb^{-1}$ given in~\cite{CMS-PAS-EXO-16-037}, we find that this upper limit corresponds to $\sigma\times \mathscr{A} \approx 40 \unit{fb}$ in our numerical simulations, which represents the current LHC limit on $\sigma\times \mathscr{A}$.~For each simplified model in Tab.~\ref{tab:benchmarks}, we express $\sigma\times \mathscr{A}$ as a function of masses and couplings as explained in Sec.~\ref{sec:jetsim}.

We also project the LHC sensitivity to $\sigma$ for an integrated luminosity of $300 \fb^{-1}$ as expected after Run 3 of the LHC as well as for the expected $3000 \fb^{-1}$ after the high luminosity LHC run. A discussion of planned detector upgrades and expected performances can be found in \Ref~\cite{CMS:2013xfa,ATL-PHYS-PUB-2014-007,CMS-PAS-FTR-13-014}, and in \Ref~\cite{Buchmueller:2014yoa} in the context of mono-jet searches for vector mediators. How exactly the sensitivity will improve for each specific search and each individual model considered here depends on the details of the detector upgrades and how much the systematic uncertainties can be improved in the future. Due to these unpredictabilities, we chose to consider two scenarios in the following: (a) the sensitivity to $\sigma$ scales with $\sqrt{L}$ and (b) linear in $L$.~Scenario (a) corresponds to a signal dominated by the statistical error on the experimental backgrounds, and (b) implies further improvements in background rejection and signal analysis (or a background-free signal region).

\begin{figure}[t]
 \begin{center}
  \includegraphics[width=\linewidth]{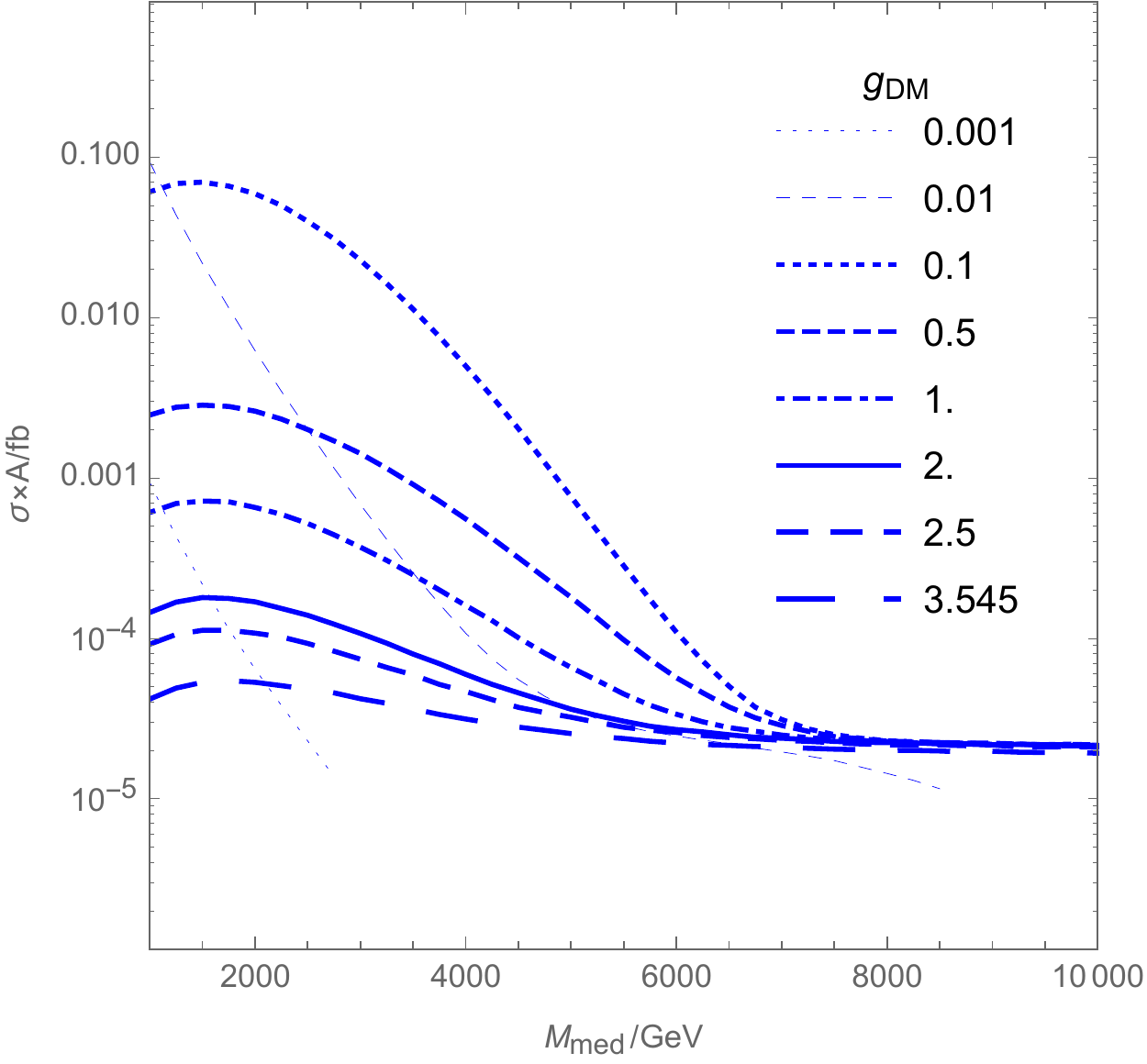}
 \end{center}
\caption{Mono-jet cross-section times detector acceptance after selection-cuts for $\hat{\mathcal{O}}_1(h_3,\lambda_3)$, with $m_\text{DM} = 50\gev$.~The mediator mass varies within $1\tev < M_\text{med} < 10 \tev$.~Different curves correspond to distinct values of $g_\text{DM}$.~Note that in the large $M_{\rm med}$ limit, all curves with not too large $\Gamma_{\rm med}$ tend to a value of $\sigma\times\mathscr{A}$ corresponding to the Effective Theory cross-section.\label{fig:opexmp}}
\end{figure}

\begin{figure*}[t]
\begin{tabular}[t]{ccc}
  \raisebox{-.8\height}{\hspace{-0.8cm}\begin{minipage}{.5\linewidth}
\includegraphics[width=\linewidth]{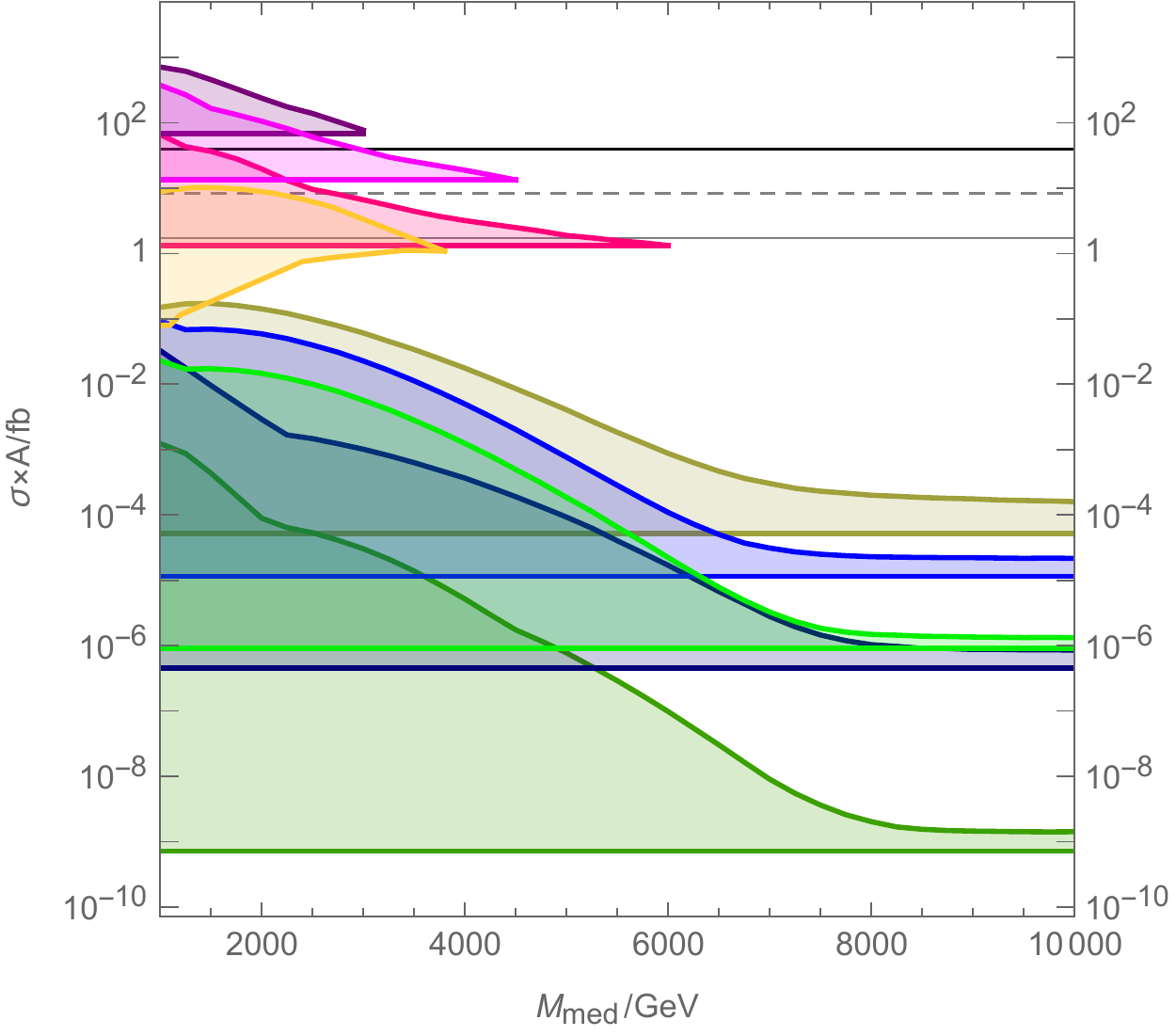}
\end{minipage}
}
&
\hspace{0.75 cm}
\begin{minipage}[t]{.15\linewidth}
\begin{footnotesize}
 \begin{tabular}[t]{ll}
  \multicolumn{2}{l}{\textbf{spin 0 DM}} \\
   \definecolor{cs1}{RGB}{60,160,0}
  \textcolor{cs1}{$\blacksquare$} & $\hat{\mathcal{O}}_1(h_1,g_1)$ \\ 
  \definecolor{cs2}{RGB}{0,240,0}
  \textcolor{cs2}{$\blacksquare$} & $\hat{\mathcal{O}}_1(h_3,g_4)$ \\ 
  \\
  \multicolumn{2}{l}{\textbf{spin $\frac{1}{2}$ DM}} \\
  \definecolor{cf1}{RGB}{0,0,120}
  \textcolor{cf1}{$\blacksquare$} & $\hat{\mathcal{O}}_1(h_1,\lambda_1)$ \\ 
  \definecolor{cf2}{RGB}{0,0,255}
  \textcolor{cf2}{$\blacksquare$} & $\hat{\mathcal{O}}_1(h_3,\lambda_3)$ \\ 
  \definecolor{cf3}{RGB}{120,0,120}
  \textcolor{cf3}{$\blacksquare$} & $\hat{\mathcal{O}}_4(h_4,\lambda_4)$ \\ 
  \definecolor{cf4}{RGB}{255,0,255}
  \textcolor{cf4}{$\blacksquare$} & $\hat{\mathcal{O}}_8(h_3,\lambda_4)$ \\ 
  \definecolor{cf5}{RGB}{255,0,120}
  \textcolor{cf5}{$\blacksquare$} & $\hat{\mathcal{O}}_{11}(h_1,\lambda_2)$ \\
\\
  \multicolumn{2}{l}{\textbf{spin 1 DM}} \\
  \definecolor{cv1}{RGB}{160,160,60}
  \textcolor{cv1}{$\blacksquare$} & $\hat{\mathcal{O}}_1(h_1,b_1)$ \\
   \definecolor{cv4}{RGB}{255,200,50}
  \textcolor{cv4}{$\blacksquare$} & $\hat{\mathcal{O}}_1(h_3,b_5)$ \\
  \end{tabular}
\end{footnotesize}
\end{minipage}

&

\begin{minipage}[t]{.15\linewidth}
\begin{footnotesize}
 \begin{tabular}[t]{ll}

  \multicolumn{2}{l}{\textbf{Limits and projections}} \\
  \definecolor{cl1}{RGB}{0,0,0}
  \textcolor{cl1}{------} & current limit \\
   \definecolor{cl1}{RGB}{180,180,180}
  \textcolor{cl1}{\,- - - } & projected sensitivity\\
  & $300~{\rm fb}^{-1}$ ($\sqrt{L}$) \\
    \definecolor{cl1}{RGB}{180,180,180}
  \textcolor{cl1}{------} & projected sensitivity\\
  & $300~{\rm fb}^{-1}$ ($L$)
 \end{tabular}
\end{footnotesize}
\end{minipage}
 \end{tabular}
\caption{XENONnT-predicted mono-jet cross-sections as a function of $M_{\rm med}$ for the models in Tab.~\ref{tab:benchmarks} compared with the LHC current limits and projected sensitivity.~We assume $\mu_{\rm S}=150$ signal events at XENONnT, $m_\text{DM} = 50\gev$ and vary the mediator mass within $1\tev < M_\text{med} < 10 \tev$.~The projected sensitivity for an integrated luminosity of $300$~fb$^{-1}$ is $\sigma\times\mathscr{A}=8.3~(1.7)$~fb, when the sensitivity scales like $\sqrt{L}$ ($L$).~For the HL-LHC run (integrated luminosity 3000 fb$^{-1}$), the projected sensitivity (not shown in the figure) is $\sigma\times\mathscr{A}=2.6~(0.2)$~fb, when the sensitivity scales like $\sqrt{L}$ ($L$).\label{fig:opall}}
\end{figure*}

\subsection{Simulation of mono-jet events}
\label{sec:jetsim}
For the simulation of mono-jet events and associated Monte Carlo integration of the production cross-section $\sigma$, we use the WHIZARD software package~\cite{Kilian:2007gr,Moretti:2001zz}.~In so doing, we implement all models in \app~\ref{app:Lagrangians} directly in WHIZARD as custom model files by extending the default Standard Model implementation.~We calculate mediator decay width and scattering cross-section for selected $2\rightarrow2$ processes involving the new particles and couplings analytically to verify our implementation.~We also use an independent implementation in MadGraph5\_aMC@NLO \cite{Alwall:2014hca} to cross-check several of our results, using custom model files generated with FeynRules \cite{Alloul:2013bka, Degrande:2014vpa, Degrande:2011ua}.~We find agreement better than $10\,\%$ between the WHIZARD and MadGraph implementations, well within the systematic error.~All calculations are performed at leading order.~However, for simplified models with (pseudo)scalar mediators we verified our results using MadGraph implementations including effective gluon-fusion vertices, which were in turn validated by one-loop calculations in MadGraph.~We compute cross-sections at the current LHC centre-of-mass energy\footnote{ When and if LHC will upgrade to $14\tev$ is unclear at this point. Moreover, this would only lead to slightly larger cross-sections and would have virtually no impact on our model comparison.} $\sqrt{s} = 13 \tev$.~We use the ``CT14lo'' pdfset interfaced via LHAPDF6 \cite{Buckley:2014ana}, which is also used to obtain the value of the strong coupling constant $\alpha_{\rm s}$ at the interaction energy scale.~Since the PDFs have to be extrapolated to the TeV range a systematic error of 10\% to 20\% cannot be avoided.~This systematic error will clearly dominate over any statistical error in our simulations.

In the calculations, we consider all processes with a pair of DM and a single (anti-)quark or gluon in the final state, $p + p \rightarrow \chi + \bar\chi + q$, $p + p \rightarrow \chi + \bar\chi + \bar q$ and $p + p \rightarrow \chi + \bar\chi + g$.~In the case of scalar or vector DM, $\chi$ is replaced by $S$ and $G^\nu$, respectively. The coloured particles will correspond to jets in the detector due to hadronization and parton showering. To allow us to compare several of the simplified models over the whole parameter space of interest in a reasonable amount of time, we will only consider event-level cross sections without showering and hadronization. We expect this to have no significant impact on our study, since reconstruction efficiency will be similar in all models considered here.~We take detector acceptance into account by imposing event-level cuts
\begin{align}
 |\eta| &< 2.5\,, &&\text{and} & \ETmiss >200 \gev\,.
\end{align}
The cut on the pseudo-rapidity $\eta$ takes the detector geometry into account, corresponding to typical cuts employed in ATLAS and CMS analyses to exclude the region close to the beam which is not well covered by trackers and calorimeters.~In particular, we use the same cuts as for deriving the limits in the previous subsection \ref{sec:limits}~\cite{CMS-PAS-EXO-16-037}.~Usually, additional cuts are imposed to avoid misidentification of jets. However, we can neglect these for our purposes, since we are only considering processes at event-level.~Furthermore, due to these cuts, we will consider only final states containing exactly one hard jet with $p_T > 100\gev$, which can be accurately calculated by a $2\rightarrow3$ matrix element. For this reason softer collinear jets from initial state radiation, which would require a more careful jet matching, do not contribute significantly to our signal and are therefore not included in our analysis.

\subsection{XENONnT-based mono-jet predictions}
\label{sec:forecasts}

\subsubsection{Predictions based on parameter inference}
In Sec.~\ref{sec:modsel}, we have shown that the detection of $\mathcal{O}(100)$ signal events at XENONnT allows to reconstruct the true value of $\mu_{\rm S}$ within a negligibly small error.~We have also argued that this holds true independently of the model underlying the signal and of the model assumed fitting the data.~Accordingly, here we assume that the detection of a signal at XENONnT implies the constraint $M_{\rm eff} = \mathscr{F}_{\rm mod}\mu_{\rm S}^{-4}$, where for a given simplified model $\mathscr{F}_{\rm mod}$ is a calculable factor.~Furthermore, here we set $\mu_{\rm S}=150$.~However, in contrast to Sec.~\ref{sec:detection}, here we calculate $\mu_S$ by integrating Eq.~(\ref{eq:rateE}) from 5 keV to 45 keV, in order to allow for a direct comparison of our results with those in~\cite{Dent:2015zpa}.~In addition, we impose that the coupling constants $g_q$ and $g_{\rm DM}$ are perturbative:~$g_q, g_\text{DM} < \sqrt{4\pi}$.~Combining the constraint on $M_{\rm eff}$ with the perturbative requirements, we obtain the regions in the $M_{\rm med} - (\sigma\times \mathscr{A})$ plane that can consistently explain the observed XENONnT signal.~While constructing such regions, for each benchmark point as listed in \reftab{tab:benchmarks} we scan over mediator masses in the range from $1$ to $10\tev$ in steps of $250\gev$, as well as over the $g_\text{DM}$ values $10^{-6}$, $10^{-5}$, $10^{-4}$, $10^{-3}$, $0.01$, $0.1$, $0.5$, $1.$, $1.5$, $2$, $2.5$, and $\sqrt{4\pi}$.~The coupling constant $g_q$ is then obtained according to \refeq{eq:Meff}, and to guarantee perturbativity we also require $g_q < \sqrt{4\pi}$.~To be consistent with the effective operator framework, the mediator width should be small compared to its mass $\Gamma_\text{med} \ll M_\text{med}$.~As a conservative upper limit we require $\Gamma_\text{med} \leq M_\text{med}$.~For parameter points fulfilling these criteria, we compute the corresponding mono-jet cross-section, accounting for the detector acceptance as described in Sec.~\ref{sec:jetsim}.~As an aside remark, we note that the limit on the decay width usually requires couplings $g_q, g_\text{DM} \ll \sqrt{4\pi}$ and is therefore more relevant.

Fig.~\ref{fig:opexmp} provides the reader with an illustrative example of the procedure described above.~It shows how the mono-jet cross-section for the $\hat{\mathcal{O}}_1(h_3,\lambda_3)$ model varies with $g_{\rm DM}$ and $M_{\rm med}$, if $M_{\rm eff} = \mathscr{F}_{\rm mod}\mu_{\rm S}^{-4}$.~In this calculation we assume $m_{\rm DM}=50$~GeV.~Taking the enveloping region of the curves for various couplings in the figure, we obtain the region in the $M_{\rm med}$ - $(\sigma \times \mathscr{A})$ plane that can consistently explain $\mu_{\rm S}=150$ in the case of the model $\hat{\mathcal{O}}_1(h_3,\lambda_3)$.

\begin{figure*}[t]
\centering
  \includegraphics[width=0.33\linewidth]{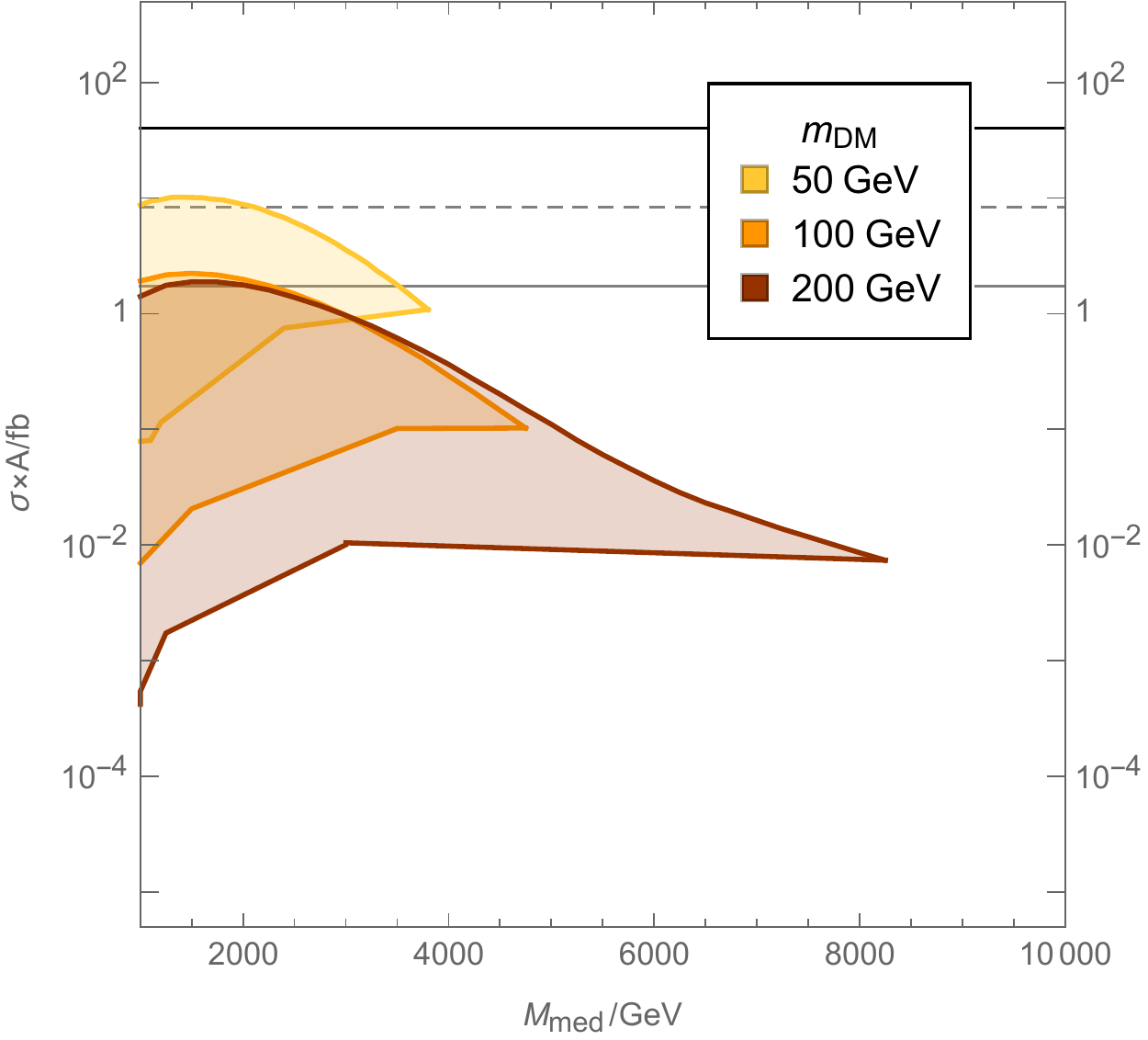}
  \includegraphics[width=0.33\linewidth]{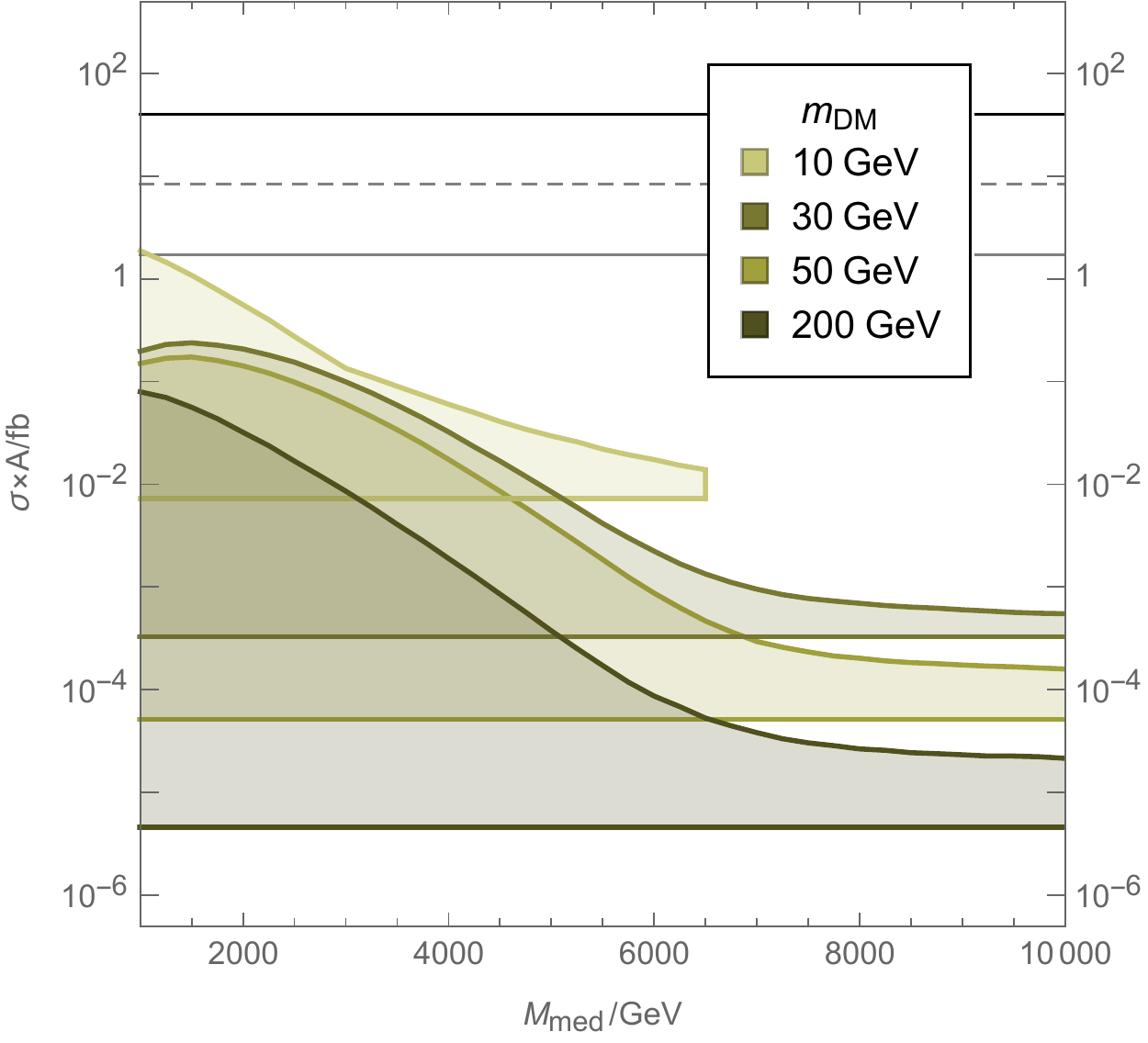}
  \includegraphics[width=0.32\linewidth]{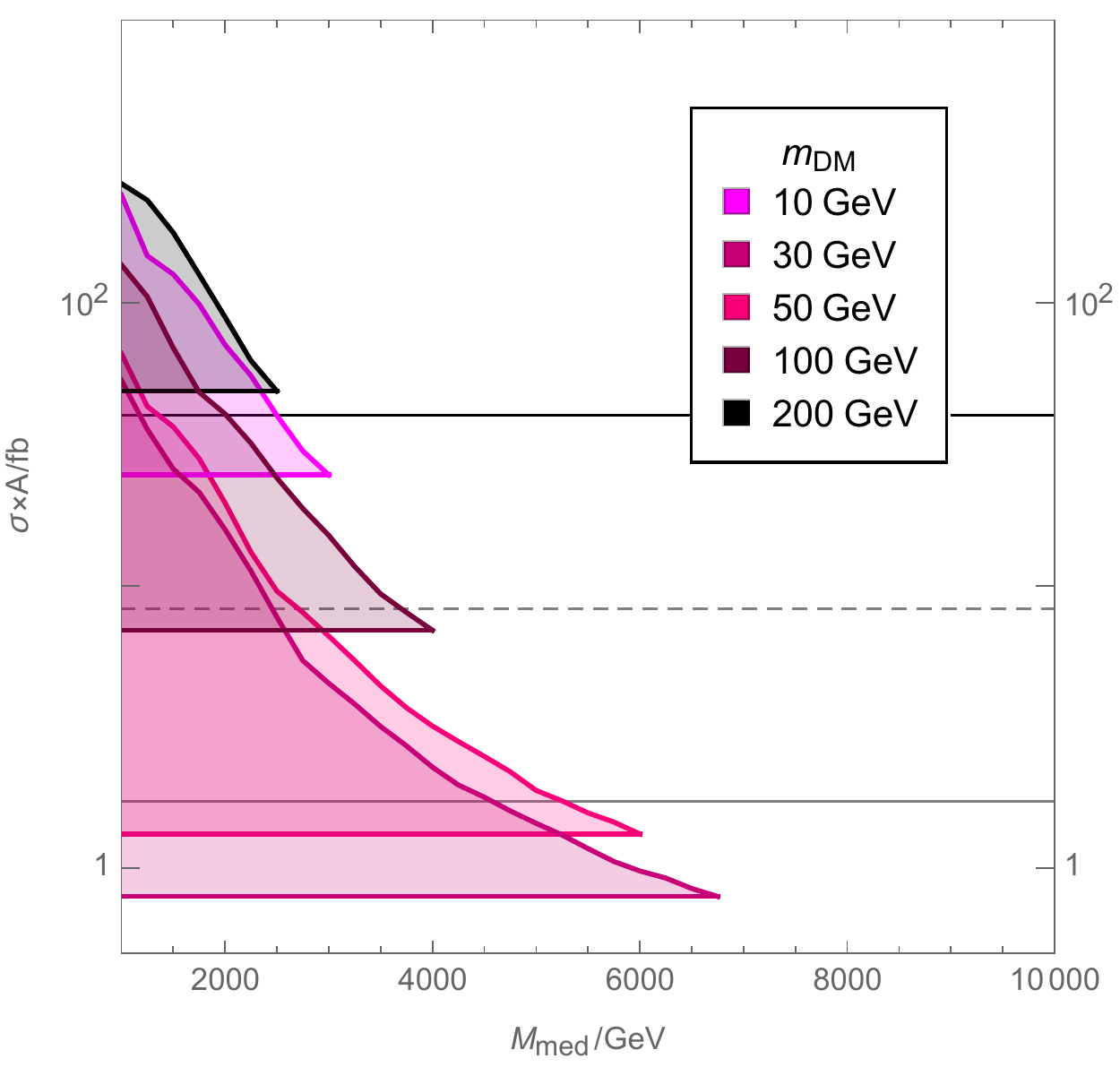}
\caption{Regions in the $M_{\rm med} - (\sigma\times \mathscr{A})$ plane that are compatible with the detection of $\mu_S=150$ signal events at XENONnT for three representative simplified models, namely $\hat{\mathcal{O}}_1(h_3,b_5)$,  $\hat{\mathcal{O}}_1(h_1,b_1)$ and $\hat{\mathcal{O}}_{11}(h_1,\lambda_2)$, and for the DM particle masses $m_{\rm DM}= 10, 30, 50, 100$ and $200$~GeV.~For model $\hat{\mathcal{O}}_1(h_1,b_1)$ the case $m_{\rm DM}=100$~GeV is omitted, since it only marginally differs from the one $m_{\rm DM}=50$~GeV.~For model $\hat{\mathcal{O}}_1(h_3,b_5)$ and for $m_{\rm DM}=10$~GeV and $m_{\rm DM}=30$~GeV, the corresponding regions in the $M_{\rm med} - (\sigma\times \mathscr{A})$ plane are empty, mainly because of the decay width constraint.\label{fig:cmp_op11h1lam2}}
\end{figure*}

Similarly, Fig.~\ref{fig:opall} shows the regions in the $M_{\rm med}$ - $(\sigma \times \mathscr{A})$ plane that can consistently explain $\mu_{\rm S}=150$ for all models in Tab.~\ref{tab:benchmarks}.~For models generating the operator $\hat{\mathcal{O}}_1$ in the non-relativistic limit, the expected mono-jet cross-sections are not within reach of the LHC Run 3.~This result is expected,  since for $M_{\rm med}\sim\mathcal{O}(1)$~TeV, relatively small values of $g_q$ and $g_{\rm DM}$ are required to obtain $\mu_{\rm S}=150$ signal events.~Small values of the coupling constants imply small mono-jet cross-sections.~Among the models generating the $\hat{\mathcal{O}}_1$ operator, model $\hat{\mathcal{O}}_1(h_3,b_5)$ predicts the largest mono-jet cross-section.~However, the predicted cross-section remains below the LHC Run 3 projected sensitivity, as long as the latter is assumed to scale like the square root of the integrated luminosity $L$.~For momentum- or velocity-dependent operators, larger coupling constants are required to obtain $\mu_{\rm S}=150$, and therefore larger mono-jet cross-sections are generically expected.~For these reasons, $\hat{\mathcal{O}}_1$ and the other operators are associated with distinct portions of the $M_{\rm med}$ - $(\sigma \times \mathscr{A})$ plane.~The only exception to this conclusion is the partial superposition between the region associated with $\hat{\mathcal{O}}_1(h_3,b_5)$ and the one corresponding to $\hat{\mathcal{O}}_{11}(h_1,\lambda_2)$.~Models not appearing in Fig.~\ref{fig:opall} are already excluded, as they would correspond to regions with $M_{\rm med}<1$~TeV and $\sigma\times \mathscr{A} \gg 40$~pb.~As for Fig.~\ref{fig:opexmp}, here we assume $m_{\rm DM}=50$~GeV.~In Fig.~\ref{fig:opall}, we also report current limits on and projected sensitivity to $\sigma \times \mathscr{A}$, estimated as described in Sec.~\ref{sec:limits} in detail.~For completeness, we compare our predictions with the projected sensitivity computed for the cases in which the sensitivity scales with $L$ linearly (e.g., as in the case of a background free experiment) and like $\sqrt{L}$ (as in the case in which the sensitivity is dominated by the statistical error on the experimental backgrounds).~However, from now onwards our conclusions will be based upon the $\sqrt{L}$ scaling, which is more conservative.~If DM is assumed to interact with quarks through one of the simplified models in Tab.~\ref{tab:benchmarks}, Fig.~\ref{fig:opall} shows where the LHC searches for mono-jet events should focus.~From this perspective, Fig.~\ref{fig:opall} represents an important guideline for the ATLAS and CMS collaborations. 

\subsubsection{Predictions based on model selection}
In Sec.~\ref{sec:modsel}, we have shown that the detection of $\mathcal{O}(100)$ signal events at XENONnT allows to discriminate featureless spectra of type $A$ from bumpy spectra of type $B$.~Combining this result with the information in Fig.~\ref{fig:opall}, we find that at the end of the LHC Run 3 only two mutually exclusive scenarios would be compatible with the detection of $\mathcal{O}(100)$ signal events at XENONnT.~In a first scenario, the spectral distribution of the $\mathcal{O}(100)$ signal events is of type $A$.~If a mono-jet signal is detected at the LHC, DM must have spin 1/2 and its interactions with nucleons must be of type $\hat{\mathcal{O}}_8$.~If a mono-jet signal is not detected at the LHC, DM-nucleon interactions must be of type $\hat{\mathcal{O}}_1$.~In a second scenario, the spectral distribution of the $\mathcal{O}(100)$ signal events is of type $B$, a mono-jet signal can be detected at the LHC, and DM must have spin 1/2 and interact with nucleons through the operator $\hat{\mathcal{O}}_{11}$.~We therefore conclude that the observation of $\mathcal{O}(100)$ signal events at XENONnT combined with the detection or the lack of detection of a mono-jet signal at the LHC Run 3 would significantly narrow the range of possible DM-nucleon interactions.~As we demonstrated above, it can also provide key information on the DM particle spin.

\section{Discussion}
\label{sec:discussion}
Here we briefly discuss the validity regime of the results found in the previous sections, and comment on possible future developments.~In Sec.~\ref{sec:forecasts}, we have presented results for $m_{\rm DM}=50$~GeV, finding that only two mutually exclusive scenarios would be compatible with the detection of $\mathcal{O}(100)$ signal events at XENONnT.~In particular, based on the hypothetical observation of $\mathcal{O}(100)$ events at XENONnT, we have shown that models generating $\hat{\mathcal{O}}_1$ are not within reach of the LHC Run 3 mono-jet searches, whereas models generating $\hat{\mathcal{O}}_8$ or $\hat{\mathcal{O}}_{11}$ can produce observable mono-jet signals.~Fig.~\ref{fig:cmp_op11h1lam2} shows that this conclusion holds true also for DM particle masses in the range $10~{\rm GeV}<m_{\rm DM}<200$~GeV.~Specifically, it shows the regions in the $M_{\rm med} - (\sigma\times \mathscr{A})$ plane that are compatible with $\mu_{\rm S}=150$ for the models $\hat{\mathcal{O}}_1(h_3,b_5)$,  $\hat{\mathcal{O}}_1(h_1,b_1)$ and $\hat{\mathcal{O}}_{11}(h_1,\lambda_2)$ for $m_{\rm DM}= 10, 30, 50, 100$ and $200$~GeV.~We find that for the two $\hat{\mathcal{O}}_1$ models, and for $M_{\rm med}\gtrsim 10$~GeV, the compatible regions remain below the projected sensitivity of the LHC Run 3, assuming that the LHC sensitivity scales like $\sqrt{L}$.~At the same time, model $\hat{\mathcal{O}}_{11}(h_1,\lambda_2)$ (and all other models different from $\hat{\mathcal{O}}_{1}$) remain testable, or partly testable, at the LHC Run 3. 

\begin{figure}[t]
 \begin{center}
  \includegraphics[width=\linewidth]{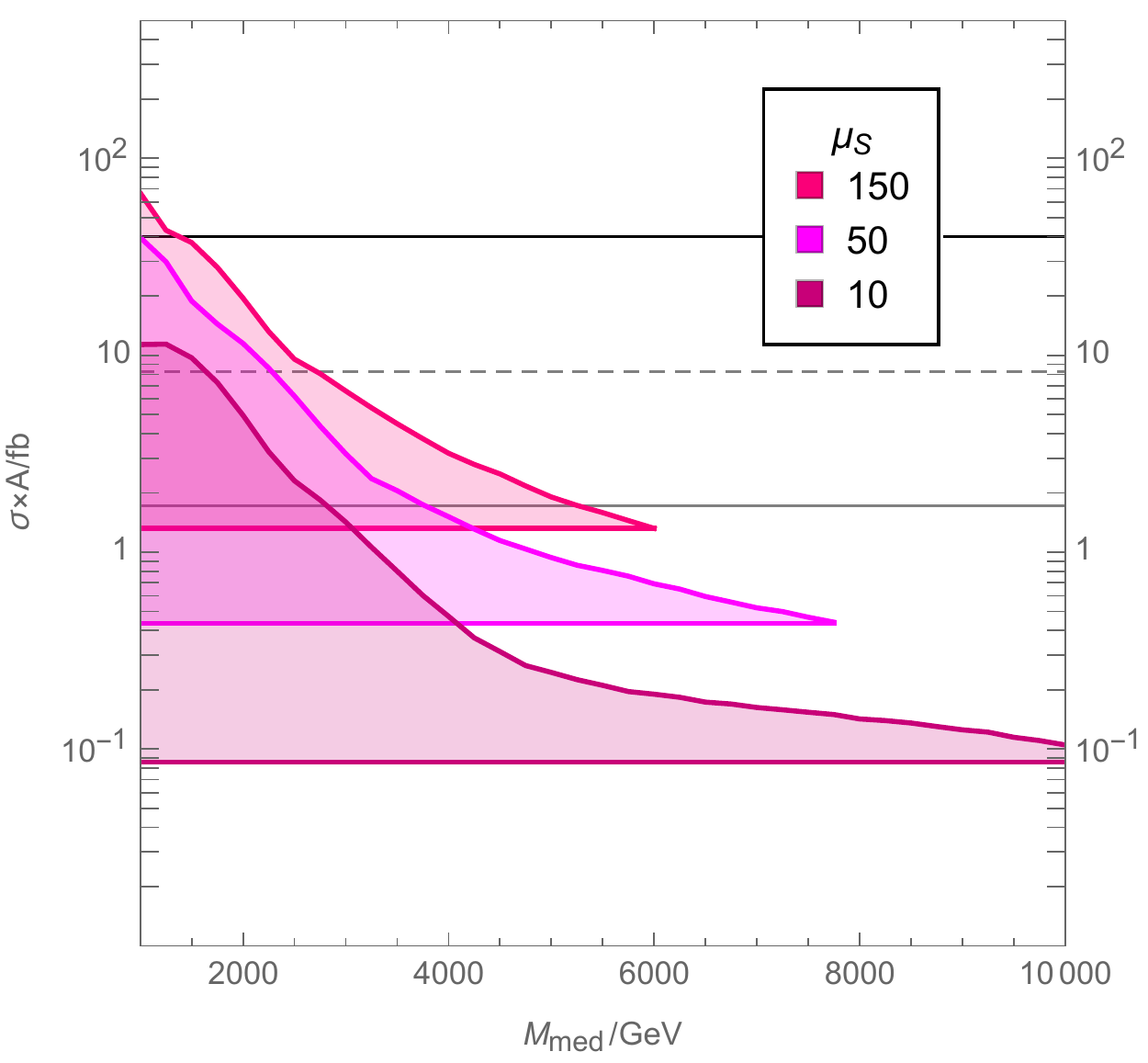}
 \end{center}
\caption{Same as for Fig.~\ref{fig:opall}, but now for different values of $\mu_S$, and for the model $\hat{\mathcal{O}}_{11}(h_1,\lambda_2)$ only.\label{fig:nus}}
\end{figure}

The results presented in the previous sections assumed $\mu_S=150$ signal events.~Here we briefly describe how our conclusions would change by decreasing $\mu_S$.~As long as $\mu_S$ is large enough such that the errors on the $y$-axis of Fig.~\ref{fig:DeltaNS} can be considered small, changing $\mu_S$ by a factor $1/x$ is equivalent to rescaling $M_{\rm eff}$ by a factor $x^{1/4}$.~This rescaling shifts the coloured regions in Fig.~\ref{fig:opall} by a factor $x$ towards smaller cross-sections.~Consequently, models which are not compatible with $\mu_S=150$, because they are already ruled out by current LHC mono-jet searches (if $\mu_S=150$), would not be in conflict with observations and remain detectable for smaller values of $\mu_S$.~For the same reasons, model $\hat{\mathcal{O}}_4(h_4,\lambda_4)$ -- arguably the most studied case of spin-dependent interaction -- would be excluded by current LHC mono-jet searches if $\mu_S=150$, but remains compatible with observations for smaller values of $\mu_S$.~The effect of rescaling the number of signal events is illustrated for the case of model $\hat{\mathcal{O}}_{11}(h_1,\lambda_2)$ and for $m_\chi=50$~GeV in Fig.~\ref{fig:nus}.

We now comment on the impact of operator mixing on our results.~When matching the simplified models in Tab.~\ref{tab:benchmarks}, which are defined at the TeV scale, with the non-relativistic operators in Tab.~\ref{tab:operators}, which are defined at the nuclear scale, one has in principle to take the evolution of the couplings between the two scales into account~\cite{Crivellin:2014qxa,DEramo:2014nmf,DEramo:2016gos}.~In most of the cases, this will lead to corrections to the scattering rates at the few percent level, which can thus be neglected to a first approximation.~There is, however, an important exception to this statement.~This is the case when the running of the coupling constants leads to a mixing between different operators.~Although coupling constants of radiatively generated operators are predicted to be a few percent of the coupling constants associated with tree-level operators~\cite{DEramo:2016gos}, the former might still generate the dominant contribution to the rate of nuclear recoils at XENONnT.~Let us illustrate this point with an example.~In the case of fermionic DM with coupling constants $h_4\neq0$ and $\lambda_3\neq0$, for instance, operator mixing introduces an interaction via $h_3$.~This new interaction will have $\hat{\mathcal{O}}_1$ in its non-relativistic limit, while one would have naively expected a (leading) contribution from $\hat{\mathcal{O}}_7$ only.~Importantly, the latter operator is proportional to $\mathbf{v}^\perp$, while $\hat{\mathcal{O}}_1$ is not.~Thus $\hat{\mathcal{O}}_1$ becomes the leading operator for direct detection experiments.~This can happen in models~\footnote{Among the models considered int \reftab{tab:benchmarks}, this is the case for scalar DM with couplings $g_4$ and $h_4$, fermionic DM with couplings $\lambda_3$ and $h_4$, and vector DM with couplings $b_5$ and $h_4$.} with an axial-vector coupling to quarks ($h_4$) mixing with a vector coupling ($h_3$), and generating the operator $\hat{\mathcal{O}}_1$.~If the induced value of $h_3$ is $\sqrt{10}$~\% of the $h_3$ value producing $\mu_{\rm S}=150$ when $M_\text{med} = 14564$~GeV and $\lambda_3=0.1$~\cite{DEramo:2016gos}, then the rescaled mediator mass needed to generate 150 events at XENONnT would be $M_\text{med} = 14564 \unit{GeV} \times \sqrt[4]{0.1} = 8190 \unit{GeV}$.~This value is much larger than $M_{\rm med}\sim\mathcal{O}(1)$~GeV, the value of  $M_{\rm med}$ needed to have a significant number of events via $\hat{\mathcal{O}}_7$ with $h_4=\lambda_3=0.1$.~Accordingly, the $\hat{\mathcal{O}}_7$ operator would give no significant contribution to the rate of signal events at XENONnT~\footnote{Considering operator evolution and mixing, however, the induced $h_3$ coupling comes from running effects and therefore is not independent on the mediator mass anymore.~Instead it depends on it logarithmically.~This dependence, however, only changes $M_{\rm eff}$ by at most an order one factor, when varying $M_{\rm med}$ from $1$ to $5\unit{TeV}$.}.~From the LHC perspective, a model with couplings $h_4$ and $\lambda_3$ would behave like model $\hat{\mathcal{O}}_1(h_3,\lambda_3)$, with an effective mass a factor of $\sqrt[4]{10}$ smaller (axial and vector quark couplings to the mediator are effectively indistinguishable at the LHC).~Consequently, for the model with couplings $h_4$ and $\lambda_3$ the predicted signal region in the $M_{\rm med} - (\sigma\times \mathscr{A})$ plane is expected to be about one order of magnitude above the one of the tree-level $\hat{\mathcal{O}}_1(h_3,\lambda_3)$ model.~This signal region would be (partly) below the projected sensitivity of the LHC Run 3.~Consequently, if running effects are not suppressed, e.g.~via non-universal DM-quark coupling or loop diagram cancellation -- as we have assumed so far -- then the lack of detection of a mono-jet signal at the LHC Run 3 would not necessarily indicate that DM interacts with nucleons through $\hat{\mathcal{O}}_1$.~Other than that, our conclusions remain unchanged.

Finally, we briefly comment on possible future developments.~The present analysis does not consider effects that might arise from a non-universal coupling of DM to quarks, from chiral effective field theory corrections to the coupling constants of the non-relativistic operators in Tab.~\ref{tab:operators}, and from mediators that are charged under the Standard Model gauge group.~We also do not consider constraints from the relic density (or discuss non-thermal production mechanisms for the simplified models considered here) and from indirect DM searches and astrophysical probes.~We leave an analysis of these aspects for future work.

\section{Conclusion}
\label{sec:conclusion}
We have developed a method to forecast the outcome of the LHC Run 3 based on the hypothetical detection of $\mathcal{O}(100)$ signal events at XENONnT.~Our method relies on the systematic classification of renormalisable single-mediator models for DM-quark interactions provided in~\cite{Dent:2015zpa}.~The proposed method consists of two stages.~In a first stage, it allows us to identify the regions in the mediator mass versus mono-jet cross-section plane that are compatible with the observation of $\mathcal{O}(100)$ events at XENONnT.~This first information is an important guideline for the experimental collaborations searching for DM at the LHC.~In a second stage, the method allows us to identify the correct (family of) DM-nucleon interaction(s) and possibly the DM particle spin based upon the observation, or the lack of observation, of a mono-jet signal at the end of the LHC Run 3.~We have applied our method to simulated XENONnT data and found that only two mutually exclusive scenarios would be compatible with the detection of $\mathcal{O}(100)$ signal events at XENONnT at the end of the LHC Run 3.~In a first scenario, the energy distribution of the simulated signal events is featureless, as for canonical spin-independent interactions described by the operator $\hat{\mathcal{O}}_1$ in Tab.~\ref{tab:operators}.~In this first scenario, if a mono-jet signal is detected at the LHC, DM must have spin 1/2 and interact with nucleons through the velocity-dependent operator $\hat{\mathcal{O}}_8$.~In contrast, if a mono-jet signal is not detected, DM must interact with nucleons through canonical spin-independent interactions.~In a second scenario, the spectral distribution of the simulated signal goes to zero in the zero momentum transfer limit.~In this second scenario, a mono-jet signal can be detected at the LHC Run 3, DM must have spin 1/2 and interact with nucleons through the momentum-dependent operator $\hat{\mathcal{O}}_{11}$.~In summary, the observation of $\mathcal{O}(100)$ signal events at XENONnT combined with the detection, or the lack of detection, of a mono-jet signal at the LHC Run 3 would significantly narrow the (as of yet broad) range of possible DM-nucleon interactions.~At the same time, it can also provide key information on the DM particle spin.

\acknowledgments We would like to thank Felix Kahlhoefer for useful insights into how to improve the LHC constraints presented here.~We also thank Andrew Cheek for pointing out some typos in Tab.~\ref{tab:coeffs}.~This work has been supported by the Knut and Alice Wallenberg Foundation (Principal Investigator: Jan Conrad;~Co-Investigators: Christian Forss\'en, Katherine Freese and Thomas Schwetz-Mangold) and is performed in the context of the Swedish Consortium for Direct Detection of dark matter (SweDCube).~K.F. acknowledges support from DoE grant DE-SC0007859 at the University of Michigan as well as support from the Michigan Center for Theoretical Physics.~K.F. and S.B. acknowledges support by the Vetenskapsr\aa det (Swedish Research Council) through contract No.~638-2013-8993 and the Oskar Klein Centre for Cosmoparticle Physics.~This investigation has also been supported by the Munich Institute for Astro- and Particle Physics (MIAPP) within the Deutsche Forschungsgemeinschaft (DFG) cluster of excellence ``Origin and Structure of the Universe''.~Finally, we would like to thank the participants of the programme ``Astro-, Particle and Nuclear Physics of Dark Matter Direct Detection'', hosted by MIAPP, for many valuable discussions. 

\appendix
\section{Lagrangians of simplified Models}
\label{app:Lagrangians}
We list here all Lagrangians for the simplified models considered in this work~\cite{Dent:2015zpa}.

\subsection{Scalar DM}
\noindent Scalar mediator:
\begin{eqnarray}
\mathcal{L}_{S\phi q} &=& \partial_\mu S^\dagger\partial^\mu S - m_S^2S^\dagger S - \frac{\lambda_S}{2}(S^\dagger S)^2 \nonumber\\
&+&\frac{1}{2}\partial_\mu\phi\partial^\mu\phi - \frac{1}{2}m_\phi^2\phi^2 -\frac{m_\phi\mu_1}{3}\phi^3-\frac{\mu_2}{4}\phi^4 \nonumber\\
&+& \ii\bar{q}\slashed{D} q - m_q \bar q q \nonumber\\
&-&g_1m_SS^\dagger S\phi -\frac{g_2}{2}S^\dagger S\phi^2-h_1\bar q q\phi-ih_2\bar{q}\gamma^5q\phi\,. \nonumber\\ 
\end{eqnarray}
Vector mediator:
\begin{eqnarray}
\mathcal{L}_{SGq} &=& \partial_{\mu}S^{\dagger}\partial^{\mu}S -m_S^2 S^{\dagger}S -\frac{\lambda_S}{2}(S^{\dagger}{S})^2  \nonumber \\
&-&\frac{1}{4}\mathcal{G}_{\mu\nu}\mathcal{G}^{\mu\nu} + \frac{1}{2}m_G^2G_{\mu}G^{\mu} -\frac{\lambda_G}{4}(G_{\mu}G^{\mu})^2 \nonumber \\
&+&i\bar{q}\slashed{D}q -m_q\bar{q}q  \nonumber\\
&-&\frac{g_3}{2}S^{\dagger}SG_{\mu}G^{\mu} -ig_4(S^{\dagger}\partial_{\mu}S-\partial_{\mu}S^{\dagger}S)G^{\mu} \nonumber\\
&-&h_3(\bar{q}\gamma_{\mu}q	)G^{\mu}-h_4(\bar{q}\gamma_{\mu}\gamma^5q)G^{\mu}\,.
\end{eqnarray}

\subsection{Fermionic DM}
\noindent Scalar mediator:
\begin{eqnarray}
\mathcal{L}_{\chi\phi q} &=& \ii\bar{\chi}\slashed{D}\chi - m_{\chi}\bar{\chi}\chi \nonumber\\
&+&\frac{1}{2}\partial_\mu\phi\partial^\mu\phi - \frac{1}{2}m_\phi^2\phi^2 -\frac{m_\phi\mu_1}{3}\phi^3-\frac{\mu_2}{4}\phi^4 \nonumber\\
&+& \ii\bar{q}\slashed{D} q - m_q \bar q q\nonumber\\	
&-&\lambda_1\phi\bar{\chi}\chi -i\lambda_2\phi\bar{\chi}\gamma^{5}\chi-h_1\phi\bar q q-ih_2\phi\bar{q}\gamma^5q\,. \nonumber\\
\end{eqnarray}
Vector mediator:
\begin{eqnarray}
\mathcal{L}_{\chi Gq} &=& \ii\bar{\chi}\slashed{D}\chi - m_\chi\bar{\chi}\chi \nonumber\\
&-&\frac{1}{4}\mathcal{G}_{\mu\nu}\mathcal{G}^{\mu\nu}+\frac{1}{2}m_{G}^2G_{\mu}G^{\mu}\nonumber\\
&+& \ii\bar{q}\slashed{D} q - m_q \bar q q \nonumber\\
&-&\lambda_{3}\bar\chi\gamma^\mu\chi G_{\mu}-\lambda_{4}\bar\chi\gamma^\mu\gamma^5\chi G_{\mu}\nonumber\\
&-&h_3\bar{q}\gamma_{\mu}qG^{\mu}-h_4\bar{q}\gamma_{\mu}\gamma^{5}qG^{\mu}\,.
\end{eqnarray}

\subsection{Vector DM}
\noindent Scalar mediator:
\begin{eqnarray}
\mathcal{L}_{X\phi q}&=&-\frac{1}{2}{\mathcal{X}}_{\mu\nu}^{\dagger}\mathcal{X}^{\mu\nu}+m_{X}^2X_{\mu}^{\dagger}X^{\mu}-\frac{\lambda_{X}}{2}(X_{\mu}^{\dagger}X^{\mu})^2 \nonumber \\
&+&\frac{1}{2}(\partial_{\mu}\phi)^2-\frac{1}{2}m_{\phi}^2\phi^2-\frac{m_\phi \mu_1}{3}\phi^3-\frac{\mu_2}{4}\phi^4 \nonumber\\
&+&\ii\bar{q}\slashed{D}q-m_{q}\bar{q}q \nonumber\\
&-&b_1m_X\phi X_{\mu}^{\dagger}X^{\mu}-\frac{b_{2}}{2}\phi^2X_{\mu}^{\dagger}X^{\mu}  \nonumber\\ 
&-&h_1\phi\bar{q}q-ih_2\phi\bar{q}\gamma^{5}q\,.  \nonumber\\ 
\end{eqnarray}
Vector mediator:
\begin{eqnarray}
\mathcal{L}_{XGq}&=& -\frac{1}{2}\mathcal{X}^{\dagger}_{\mu\nu}\mathcal{X}^{\mu\nu}+m_{X}^2X^{\dagger}_{\mu}X^{\mu}-\frac{\lambda_{X}}{2}(X_{\mu}^{\dagger}X^{\mu})^2 \nonumber\\
&-&\frac{1}{4}\mathcal{G}_{\mu\nu}\mathcal{G}^{\mu\nu}+\frac{1}{2}m_{G}^2G_\mu^2-\frac{\lambda_G}{4}(G_\mu G^\mu)^2\nonumber \\
&+&i\bar{q}\slashed{D}q-m_{q}\bar{q}q\nonumber\\
&-&\frac{b_3}{2}G_\mu^2(X^{\dagger}_\nu X^{\nu}) -\frac{b_{4}}{2}(G^{\mu}G^{\nu})(X^{\dagger}_{\mu}X_{\nu}) \nonumber\\ 
&-&\left[ib_{5}X_{\nu}^{\dagger}\partial_{\mu}X^{\nu}G^\mu+b_{6}X_{\mu}^{\dagger}\partial^\mu X_{\nu}G^{\nu}\right.\nonumber\\ 
&+& \left.b_{7}\epsilon_{\mu\nu\rho\sigma}(X^{\dagger\mu}\partial^{\nu}X^{\rho})G^{\sigma} +h.c.\right]\nonumber\\
&-&h_3G_\mu\bar{q}\gamma^\mu q - h_4 G_\mu\bar{q}\gamma^\mu\gamma^{5}q\,. 
\end{eqnarray}

%

\end{document}